\documentclass[aip,amsmath,amssymb]{revtex4-2}

\usepackage{graphicx}
\usepackage{dcolumn}
\usepackage{bm}

\usepackage[utf8]{inputenc}
\usepackage[T1]{fontenc}
\usepackage{mathptmx}
\usepackage{etoolbox}
\usepackage{xcolor}

\makeatletter
\def\@email#1#2{%
 \endgroup
 \patchcmd{\titleblock@produce}
  {\frontmatter@RRAPformat}
  {\frontmatter@RRAPformat{\produce@RRAP{*#1\href{mailto:#2}{#2}}}\frontmatter@RRAPformat}
  {}{}
}%
\makeatother
\begin{document}

\title[Temperature limits of quantum dynamics]{Temperature limits of the transitional quantum dynamics in qubit clusters exposed to the ac field}

\author{S.~E.~Shafraniuk}
\affiliation{Tegri LLC, 558 Michigan Ave, Evanston, 60202, IL, USA}
\pacs{DOI: 10.1109}

\date{\today}



\begin{abstract} 
Extending the temperature limits of quantum coherence in the system representing a chain of coupled two-level systems (TLS) exposed to an electromagnetic field is complicated due to the adverse influence of noise. Such a system frequently serves as a basic element of various quantum devices. In the steady state, the quantum coherence in TLS is merely destroyed by noise, which intensifies as the temperature increases. The behavior is complicated when the external field is applied modulating also the noise. In this work, using the computerized model, we study the temperature limits of the transitional quantum dynamics in the all-electrically controlled graphene single-TLS and three-TLS devices exposed to the electromagnetic field. We analyze how the external ac field changes the state of the system and observe that it not only influences the coherent transport there but modifies the effect of noise. The conducted numerical experiments determine the conditions provided the quantum coherence in QC may be much prolonged even above the ambient room temperature which can improve the performance of various quantum devices.
\end{abstract}

\maketitle

\section{Introduction}\label{Intro}
Raising the temperature limits of the functionality of various terahertz sensors, analyzers, lasers, and quantum computing devices is technically challenging because the energy of photons is small and thus the devices are vulnerable to thermal fluctuations \cite{Schoen-2020,Shafr-PRB-2019,Shafr-AQT-2020,Asada-2021,Shafr-PRB-2023,Shafr-NL-2012,Shafr-JP-2011,Shafr-EPJ-2011,Shafr-JP-2008,Shafr-PRB-2008,Shafr-PRB-2007,Grifoni-1998,Shevchenko,Shafr-Graph-Book-2015}. Despite the challenges, the nowadays demands to physical devices require improving the high-temperature performance, reduced noises,  increased detectivity, tuneability, and multispectral functionality.

Therefore, a lot of attention nowadays is paid to microscopic mechanisms responsible for the performance of various quantum wells and quantum dot devices serving as qubits, ac field sensors, and lasers, whose physics and fabrication technology are intensively developed nowadays \cite{Shafr-Graph-Book-2015,Koppens}. An increased interest is directed toward all-electrically controlled solid-state quantum devices involving low-dimensional structures, and nanotechnology \cite{Shafr-Graph-Book-2015,Koppens}. The novel solutions contemplate efforts to reduce the influence of noise by exploiting the quantum phenomena such as quantization, intrinsic spectral narrowing \cite{Shafr-AQT-2020,Asada-2021,Shafr-PRB-2023}, and Rabi flops \cite{Schoen-2020}. A well-known technique to mitigate high loss is the strong noiseless amplification before detection and the phase-sensitive amplification of squeezed states \cite{Frascella-2021,Devoret,Kwan-Squeeze}.

\begin{figure}
\includegraphics[width=125 mm]{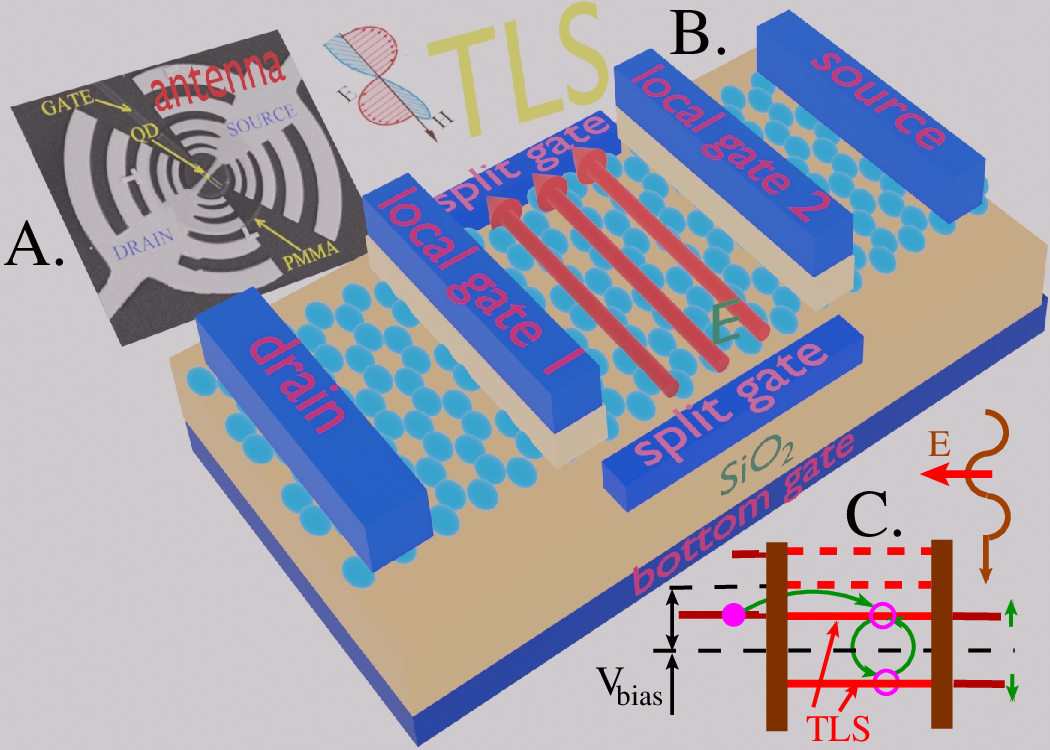} 
\caption{ {A.~A bow‐tie antenna with the QC device in the center. B.~The electrically controlled single-TLS device formed on a graphene stripe with zigzag atomic edges (ZZ-stripe). The bias voltage $V_{\rm bias} $ applied to the source and drain electrodes induces electron interlevel transitions and causes interdot hoppings (see green arrows in the energy diagram C)  \cite{Shafr-Floq}. The split gates create a transversal electric field ${\bf E}$ causing the Stark splitting $\Delta $ of the zero-energy level into the two voltage-controlled TLS levels marked by green "$\uparrow $" and "$\downarrow $". C.~The energy diagram shows the ac field-induced Floquet processes discussed in Ref.~\cite{Shafr-Floq}. The dashed lines mark the Floquet pseudoenergies defined in Sec.~\ref{sec-Fl-Mark}. The bottom gate controls the mean value of the electrochemical potential $\mu $ in the ZZ-stripe while local gates tune the height of the interdot chiral barrier $V_{\rm B}$.}}
\label{Fig_1}
\end{figure}

In this work, using numerical experiments, we address the temperature limits of the transitional quantum dynamics in the qubit cluster (QC) comprising one or several quantum dots exposed to the electromagnetic field. The QC device can serve as a basic element of quantum computing circuits or a quantum sensor. The computerized model of a real QC device harbors a two-level system (TLS) formed in either a single quantum dot (see Fig.~\ref{Fig_2}) or in a chain of them. Although the model applies to QC of any geometry and material, here we consider dots by size $\sim 2-20$~nm, which are formed on graphene nanoribbon as sketched in Figs.~\ref{Fig_2} and \ref{Fig_9}. On the one hand, in the steady state, when the ac field is off, the quantum coherence QC is improved owing to the conventional~\cite{Asada-2021} or intrinsic spectral narrowing~\cite{Shafr-AQT-2020,Shafr-PRB-2023}. On the other hand, the intrinsic noise influence on quantum coherence in the steady state is adverse and intensifies as the temperature increases. The behavior becomes more complicated when the external field is on. We examine how the external ac field influences the coherent transport in QC and modifies the effect of noise. The problems to study along with changes in the influence of noise are temperature fluctuations, spectral line broadening, decoherence, dephasing, and resilience of quantum effects. Improving the operational performance of quantum devices is based on optimizing the transient quantum dynamics. We aim to establish the conditions when the quantum coherence in QC remains sufficiently long even at increased temperatures which would result in better functionality of quantum dot devices.

In the next Sec.~\ref{methods} we outline the approach used for finding the transitional quantum dynamics of TLS chains under the influence of ac field and computing the spectral density of noise in graphene stripe. In Sec.~\ref{noise} we discuss the spectral density of noise $S(\omega, T)$ and formulate a simple model form of it to be used in calculations of the quantum system characteristics of interest. Next, in Sec.~\ref{single-TLS} we formulate a numerical model to study the temperature effect of noise in the single TLS under the influence of ac field. By solving the Lindblad master equation and using the Floquet-Markov formalism \cite{Grifoni-1998,Shevchenko,Creffield-2003,Breuer,Haug,Nielsen} (see also Sec.~ref{sec-Fl-Mark}), we compute the expectation values such as occupation probabilities of quantized levels, destruction of Rabi flops by noise at higher temperatures, and evolution of excitation spectrum in intensive ac field. In Sec.~\ref{three-dot} we address the three-TLS device, whose energy spectrum experiences either the conventional~\cite{Asada-2021} or intrinsic spectral narrowing~\cite{Shafr-AQT-2020,Shafr-PRB-2023} resulting in considerable shrinking the quantized energy level width by 4-6 orders of magnitude. The mentioned phenomenon leads to the reduction of the adverse noise influence by a few orders of magnitude causing the respective prolongation and resilience of the quantum coherence. In Sec.~\ref{discuss} we discuss the obtained results and summarize the effect of ac field on the temperature changes in the transitional quantum dynamics in qubit clusters. The conclusions are listed in Sec.~\ref{conclude}. A brief description of methods such as the Lindblad master equation  \cite{Grifoni-1998,Shevchenko,Creffield-2003,Breuer,Haug,Nielsen}, and Floquet-Markov formalism described in Secs.~\ref{Linbl} and \ref{sec-Fl-Mark}.

\section{Methods}\label{methods}

There are two kinds of quantum systems. One kind is open systems interact with a larger environment and closed systems that do not. The other systems are closed systems, whose state can be described by a state vector, although when there is entanglement a density matrix may be needed instead. Below we will model an open system or an ensemble of systems using the density matrix formalism.
The dynamics of an open system given its initial state, a time-dependent Hamiltonian,
a list of operators through which the system couples to its environment and
a list of corresponding spectral-density functions that describes the
environment are obtained by solving the Lindblad equation (see Sec.~\ref{e-ph}) and the Floquet-Markov master equation (see Secs.~\ref{Linbl} and \ref{sec-Fl-Mark}) by using QuTiP \cite{QuTiP1,QuTiP2}. The QuTiP solvers characterize the environment with a decay time $\tau_{\rm dc}$, by extracting the strength of the coupling to the environment from the noise spectral-density functions and the instantaneous Hamiltonian parameters. The numerical model involves several assumptions to derive the Lindblad master equation from physical arguments. Although the mentioned assumptions limit the applicability of the Lindblad master equation they are physically meaningful and essentially simplify the numeric calculations by using the standard solvers in QuTiP \cite{QuTiP1,QuTiP2} as described below. The assumptions are as follows. ({\i }) Separability, which means that at $t=0$ there are no correlations between the system and its environment. The full density matrix then becomes a tensor product $\rho _{\mathrm{tot}}^{I}\left( 0\right) =\rho ^{I}\left( 0\right) \otimes \rho _{\mathrm{env}}^{I}\left( 0\right) $. ({\it ii}) Born approximation that requires that ({\it a}) the interaction with the system does not change the state of the environment; ({\it b}) throughout the evolution, the system and the environment remain separated. The above remains true if the interaction is weak, and if the environment is much larger than the system. Summarizing, $\rho _{\mathrm{tot}}\left(t\right) =\rho \left( t\right) \otimes \rho _{\mathrm{env}}$. ({\it iii}) under the Markov approximation, one assumes that the timescale of decay for the environment $\tau _{\mathrm{env}}$ is much shorter than the smallest timescale of the system dynamics $\tau _{\mathrm{sys}}\gg \tau _{\mathrm{env}}$. The Markov approximation is often quoted as a  \textquotedblleft short-memory environment\textquotedblright\ as it requires that environmental correlation functions decay on a time-scale fast compared to those of the system. ({\it iv}) Secular RWA approximation (see Sec.~\ref{RWA}) assumes that all fast-rotating terms in the interaction picture can be neglected. One also neglects terms that lead to a little renormalization of the system energy levels. That approximation is not mandatory for all master-equation formalisms (e.g., the Block-Redfield master equation), but it is required for arriving at the Lindblad form which is used in QuTiP \cite{QuTiP1,QuTiP2}. For the sake of simplicity, we consider only the monochromatic ac field. Besides, we assume that the quantum coherence spreads over several quantum dots, as was predicted for graphene QC in Ref.~\cite{Shafr-AQT-2020,Shafr-PRB-2023}, thus there is no dc bias between the quantum dots connected in sequence. The dc bias, when applied, develops only on the edges of the multiple quantum dot sequence but not inside of it. 

The numerical model below applies to a variety of solid-state low-dimensional quantum dot devices. Here we consider an illustrative example such as the QC device based on the graphene nanoribbons with zigzag atomic edges (ZZ-stripes), which have numerous benefits~\cite{Shafr-AQT-2020,Shafr-PRB-2023}. ({\it i}) The edge states are robust against the electron scattering on lattice imperfections and phonons because they are topologically protected~\cite{ZZ-stripe-topolog-insul-2011,Arabs}. ({\it ii}) There are two well-defined LS levels in the electron spectrum in ZZ-stripes. The respective level spacing is readily controlled by the split gates. This simplifies manipulations by the quantized states and makes the multi-qubit operations feasible. ({\it iii}) The DOS peaks at the LS energies are very sharp and therefore, the electron-phonon scattering is weakened. Thus the quantum coherence remains preserved even at elevated temperatures, opening the way to a flexible all-electrical control in the graphene quantum dot systems. The level width $\Gamma $ is diminished by several orders of magnitude owing to the intrinsic spectral narrowing of the energy level singularities. Then, by reducing the coupling of qubits to a noisy environment and by eliminating the inelastic electron-phonon scattering one can prolong the coherence time significantly above $T\sim 300$~K. This promises a stable operation of respective multi-qubit circuits up to room temperature.

\section{Spectral density of the noise}\label{noise}

The limited coherence time of the two-level systems (TLS)~\cite{Devoret,Ithier} represents a fundamental roadblock on the way to feasible detecting and quantum computing applications. The TLS couple to and thus dissipate information into the noisy environment. Longitudinal coupling describes (pure) dephasing, while transverse coupling is responsible for relaxation.

The noise in a solid-state TLS device originates from a variety of uncorrelated sources. They include Johnson-Nyquist (JN), photon, amplifier, load, and excessive noise. The JN noise is generated by the thermal agitation of the charge carriers (usually the electrons) at equilibrium inside an electrical conductor with resistance $R$, which happens regardless of any applied voltage. Below, for the sake of simplicity, we assume the following. The one-sided power spectral density of JN is $S_{\mathrm{JN}} = 4k_{B}TR$, where $T$ is the temperature. In the last expression, the effective $R$ is rather low and in our model, we include $S_{\mathrm{JN}} $ into the additional term $S_{\rm others}\left( \omega, T \right) $ accounting also for respective contributions from other noise sources such as photon, amplifier, load, and excessive noise. the total spectral density as a small addition. The photon noise is the randomness in signal associated with photons arriving at the TLS device. Below, for the sake of simplicity, we neglect the photon noise by considering the purely harmonic signal with no randomness. The excess electrical noise involves two types, flicker (\textquotedblleft $1/f$\textquotedblright ) noise and contact shot noise, which may become significant at low frequencies. Both noises are mimicked here by simple analytical expressions. We assume that the noise is mostly generated in the substrate and the external electrodes.  The protection from external noise was discussed in Ref.~\cite{Mayle-JP-2015}, where a bi-metal multilayer system was proposed. The bi-metal multilayer filters the heat flow by separating the electron and phonon components one from another. The multilayer minimizes the phonon component of the heat flow while retaining the electronic part. The method~\cite{Mayle-JP-2015} allows for improving the overall performance of the electronic nano-circuits due to minimizing the energy dissipation.  Such idea~\cite{Mayle-JP-2015} can be exploited to reduce the external noise influence in the devices depicted in Figs.~\ref{Fig_1} and~\ref{Fig_9}.

We consider the effect of noise by mapping the noisy quantum simulator to a system of fermions coupled to a bath, as for the electron-phonon coupling (see, e.g., Ref. \cite{Zanker}. To understand the effect of decoherence and to model the transient evolution of TLS after an initialization into a non-thermal state, we start from a single TLS and only then we examine the chain of three coupled TLS with the dissipative decay due to a bosonic bath and decay due to TLS. The external noise acting on the open quantum system causes the incoherence consisting of the "longitudinal" and "transverse" processes characterized by the relaxation rate $\Gamma _{1}$ and dephasing rate $\Gamma _{2}$ respectively. In the adiabatic limit, we consider a TLS subjected to decay due to a bath of harmonic oscillators characterized by its power spectral density $S(\omega ) = J_{i}(\omega )\coth (\omega /2T)$, where $J_{i}(\omega ) $ is the electron-phonon spectral function. We assume that the TLS interacts linearly with the displacement of the oscillators and the interaction of the quantum states with the noise is identical to a local interaction between electrons and phonons. The above is used for mimicking the spectral density of graphene. 
The noise generated in course of electron-phonon collisions is discussed in Sec.~1 of Appendix.

\begin{figure}
\includegraphics[width=125 mm]{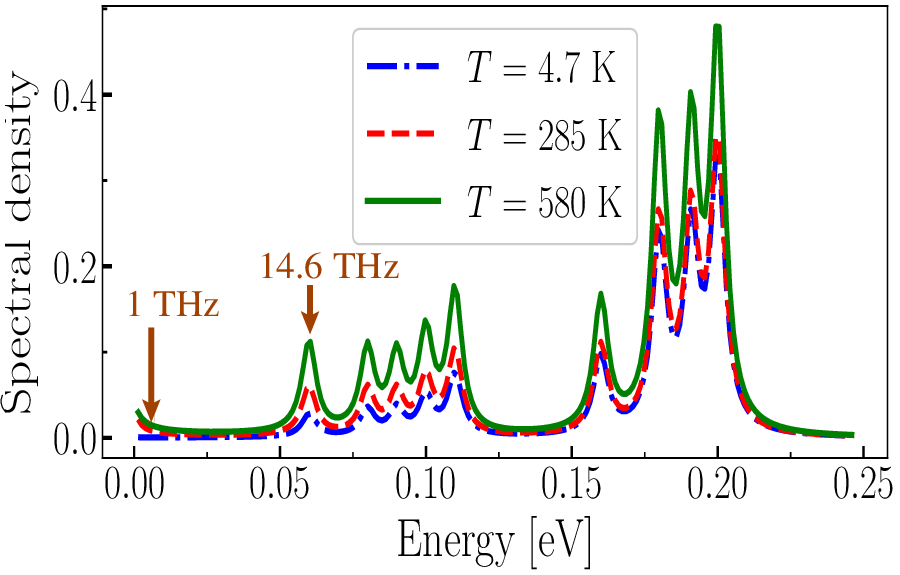} 
\caption{The energy dependence of the noise spectral density $S\left( \omega \right) $ in the graphene stripe for different temperatures. To emphasize the various temperature dependence of the noise intensity at different energies,  we denote the frequencies 1~THz and 14.6~THz by arrows. Below we will see that the mentioned difference is important for establishing the temperature limits of the transitional quantum dynamics in qubit clusters exposed to the ac field.}
\label{Fig_2}
\end{figure}

When solving the Floquet-Markov equation for the TLS device exposed to the ac field one considers the following. The alternating field not only affects the electron states but also modifies the effect of noise. Therefore, it is technically convenient to reduce all the noise sources in the system to a single effective source of noise described by some model shape of the spectral density that summarizes all the partial components. In the toy model, the spectral density of the noise is mimicked as%
\begin{equation}
S\left( \omega, T \right) =\gamma _{\mathrm{noise}}\left [ \sum_n \frac{\coth \left( \omega /T\right) }{\left( \omega -\omega _{n}\right) ^{2}+ (\gamma _{\mathrm{ph}%
}^{\left( n\right) })^2} + S_{\rm others}\left( \omega, T \right)  \right ]                                         \label{Snoise}
\end{equation}%
where $T$ is temperature, $\omega _{n}$ are the phonon branches (e.g., for the graphene monoatomic layer $n=9$), $\gamma _{\mathrm{noise}}$ is the noise intensity, $\gamma _{\mathrm{ph}}^{\left( n\right) }$ is the width of $n $-th phonon branch. One then appropriately selects parameters of the electron-phonon interaction, positions, and widths of phonon branches in  Eq.~(\ref{Snoise}), and also adds the term $S_{\rm others}\left( \omega, T \right) $ accounting for respective contributions from other noise sources such as Johnson-Nyquist (JN), photon, amplifier, load, and excessive noise. As is evident from Fig.~\ref{Fig_2}, the resulting $S\left( \omega, T \right) $ is frequency- and temperature-dependent. One must be advised that the shape~(\ref{Snoise}) of $S\left( \omega, T \right) $ is just illustrative. The exact shape of the noise spectral function depends on a variety of microscopic and geometrical factors, such as the stripe thickness, substrate, heat/noise protection shields, the overall device geometry, etc. We will not concentrate here on the aforementioned details, since our wish is to find general recipes to improve the high-temperature functionality of the TLS devices. The most important feature in Fig.~\ref{Fig_2}, is that the temperature dependence of $S\left( \omega, T \right) $ is visible only for frequencies coinciding with the peaks in $S\left( \omega, T \right) $. Remarkably, in the flat portions of $S\left( \omega, T \right) $  in Fig.~\ref{Fig_2} the temperature dependence is negligible. The mentioned factors impact the modeling results reported below.

Below we discuss the numerical results obtained using the QuTiP solvers \cite{QuTiP1,QuTiP2} allowing computing the Floquet modes and quasi energies [see Eqs.~(\ref{Fl-mod}) and (\ref{Eig})], Floquet state decomposition, etc., given a time-dependent Hamiltonian (\ref{Sht}) on the callback format, list-string format, and list-callback format (see Ref.~\cite{Shafr-Floq} and Secs.~\ref{Linbl} and \ref{sec-Fl-Mark}). The Floquet modes at $t=0$ corresponding to the Hamiltonian of the system are  calculated using the \emph{qutip.floquet.floquet\_modes} function, which returns lists containing the Floquet modes and the quasienergies. 

\section{Numerical experiments}\label{experiment}

\subsection{Single TLS exposed to the ac field}\label{single-TLS}
An instructive numerical model is the single-TLS device exposed to the ac field, as depicted in Fig.~\ref{Fig_1}. The transient dynamics of such a device are described by the time-dependent Hamiltonian 
\begin{equation}
{\cal H}(t)=-\frac{\delta }{2}\sigma _{x}-\frac{\epsilon _{0}}{2}\sigma _{z}+\frac{A}{2}\sigma
_{x}\sin \omega t
\end{equation}%
where $\delta $ is the interlevel coupling energy, $\epsilon _{0}$ is the
quantized level spacing, $A$ is the ac field's amplitude, $\omega $ is
the angular frequency, and $\sigma _{i}$ ($i=x,y,z$) are Pauli matrices.

\begin{figure}
\includegraphics[width=125 mm]{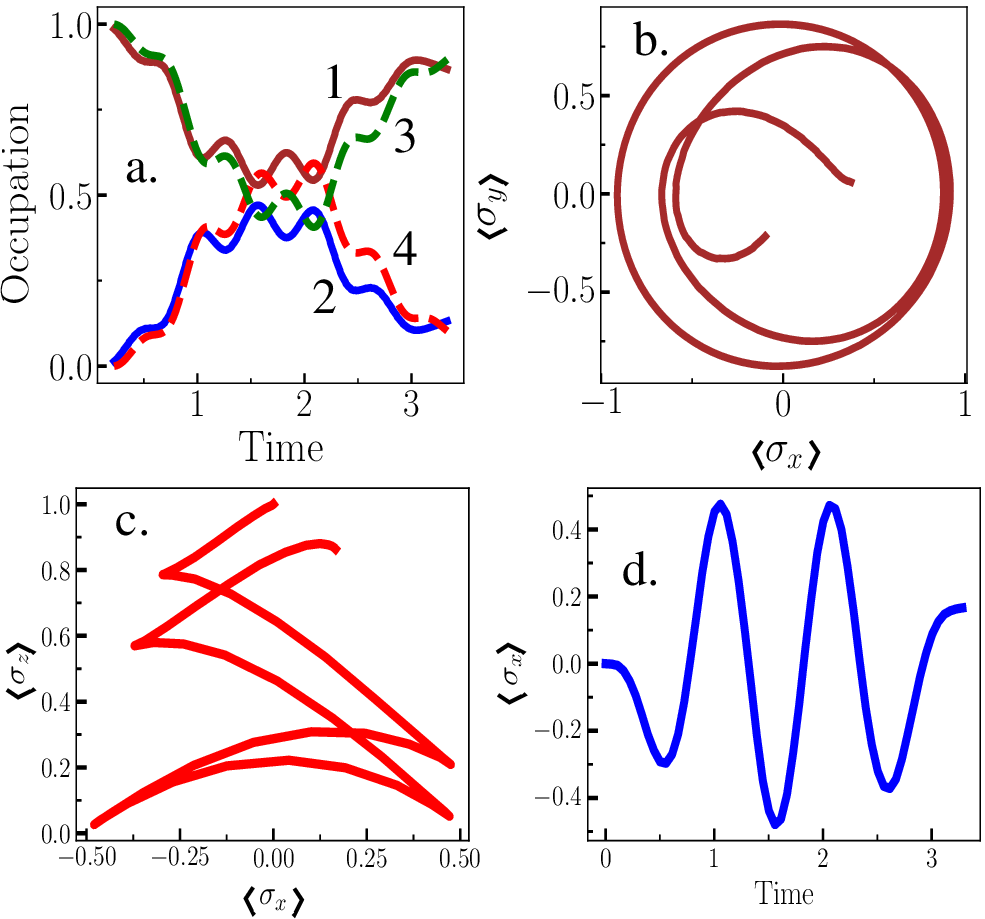} 
\caption{ {(a)~The one-period time dependence of the occupation probabilities for lower ($P_{0}^{\rm Fl} $, curve~1) and upper ($P_{1}^{\rm Fl}$, curve~2) energy levels spaced by $\epsilon _{0}=1$ QU computed using the Floquet-Markov formalism (see Secs.~\ref{Linbl} and \ref{sec-Fl-Mark}) for one period of time in the single-TLS device (solid curves). Respectively, the dashed curves $P_{0,1}^{\rm L}$ (curves 3, 4) are obtained by solving the Lindblad equation. (b)~Parametric plot of expectation values $%
\left\langle \sigma _{y}\right\rangle $ vs $\left\langle \sigma_{x}\right\rangle $.
(c)~Parametric plot of expectation values $\left\langle \sigma _{z}\right\rangle $ vs $\left\langle \sigma _{x}\right\rangle $; (d)~Single period of
expectation value $\left\langle \sigma _{x}\left( t\right) \right\rangle $.%
}}
\label{Fig_3}
\end{figure}

To obtain the Floquet quasienergies according to Secs.~\ref{Linbl} and \ref{sec-Fl-Mark}, the numerical technique described in Refs.~\cite{QuTiP1,QuTiP2} was used.  In Sec.~\ref{sec-Fl-Mark}, the Floquet modes $\varepsilon_{\alpha }$ [see Eqs.~(\ref{Fl-mod}) and (\ref{Eig})] are eigenstates of the one-period propagator for the time-dependent Schr\"{o}dinger equation (\ref{Sht}). By numerically calculating $U\left( T+t,t\right) $ and diagonalizing it, we evaluate the unitary evolution operator for one period of the field $U\left( T,0\right) $ and obtain its eigenvalues, which are related to the quasienergies via  $\epsilon _{\alpha }=-\hbar \arg \left( \eta _{\alpha }\right) /T$. By calculating and diagonalizing $U\left( T,0\right) $, this gives $\Phi _{\alpha }\left( 0\right) $. 
Next, using the wave function propagator $U\left( t,0\right) \Psi _{\alpha }\left(0\right) $ for an arbitrary time moment $t$, the Floquet modes are computed by propagating $\Phi _{\alpha }\left( 0\right) $ to $\Phi _{\alpha }\left( t\right) $. For the Floquet modes this gives $U\left( t,0\right) \Phi _{\alpha }\left( 0\right) =\exp \left( -i\epsilon _{\alpha }t\right) \Phi _{\alpha }\left( t\right)$, so that $\Phi _{\alpha }\left( t\right) =\exp \left( i\epsilon _{\alpha }t\right) U\left( t,0\right) \Phi _{\alpha }\left( 0\right) $. Because $\Phi_{\alpha }\left( t\right) $ is periodic, we only need to evaluate it only for one period of field $t\in \lbrack 0,T] $.  For arbitrary long time $t$, using $\Phi _{\alpha }\left( t\in \lbrack 0,T]\right) $, we directly evaluate $\Phi _{\alpha }\left( t\right) $, $\Psi _{\alpha }\left( t\right) $, and $\Psi \left( t\right) $.

\begin{figure}
\includegraphics[width=125 mm]{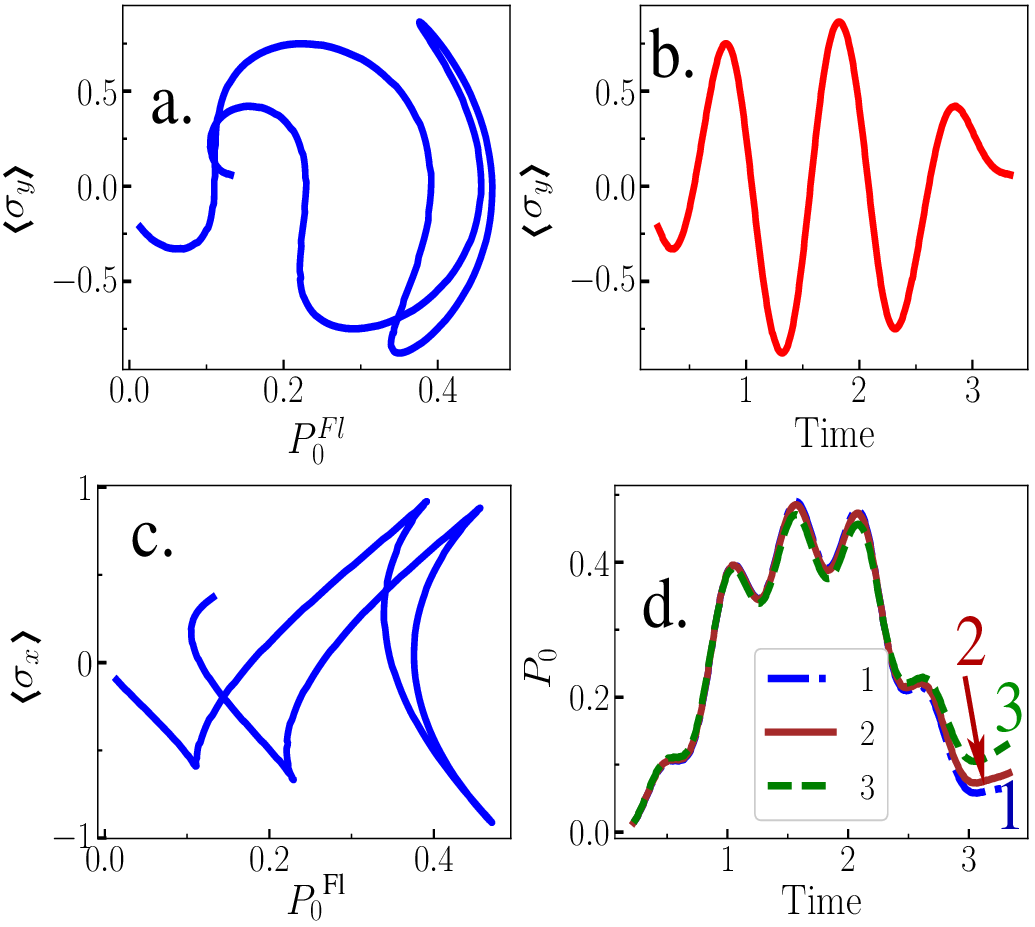} 
\caption{ {(a) Parametric plot of expectation values $\left\langle \sigma _{y}\right\rangle $ vs the occupation probability of lower level $P_{0}^{\rm Fl}$. (b) Single period of expectation value $\left\langle \sigma _{y}\left( t\right) \right\rangle $.
(c) Parametric plot of expectation values $\left\langle \sigma _{x}\right\rangle $ vs the occupation probability of lower level $P_{0}^{\rm Fl}$; (d) Single period of the occupation probability of lower level $P_{0}^{\rm Fl}$ for three temperatures $T = 33$~K, $280$~K, and $570$~K (curves 1-3 respectively).
}}
\label{Fig_4}
\end{figure}

A driven system that is interacting with its environment is not necessarily well described by the standard Lindblad master equation~(4) (see in Sec.~1) since its dissipation process becomes time-dependent due to the driving. In such cases, a rigorous approach would be to take the driving into account when deriving the master equation. This can be done in many different ways, but one-way common approach is to derive the master equation in the Floquet basis. That approach results in the so-called Floquet-Markov master equation (see Ref.~\cite{Grifoni-1998} for details). Using this method to obtain the quasienergies, a bisection algorithm is used to find the location of the quasi-energy crossings to a high degree of accuracy. Besides, the dynamical behavior of the system was examined directly by integrating it over long periods, with the particle initially located in the left quantum dot. 

\begin{figure}
\includegraphics[width=125 mm]{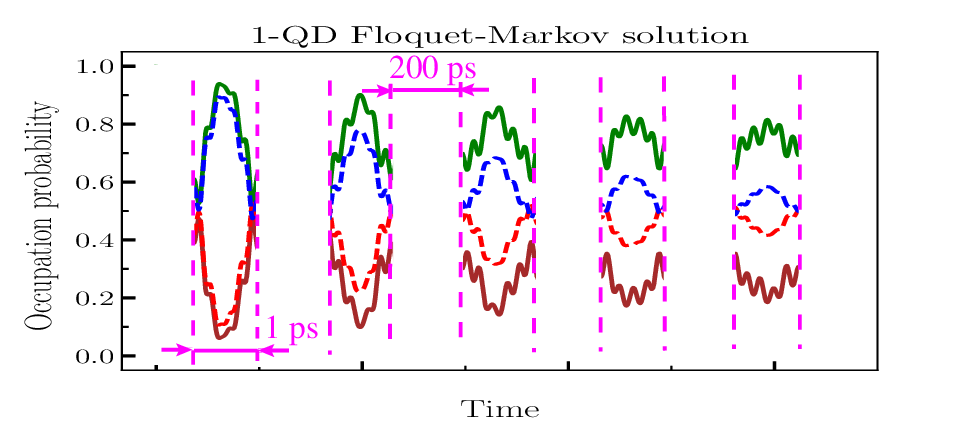} 
\caption{ { The occupation probabilities $P_{0}^{\rm Fl}$ (solid green) and $P_{1}^{\rm Fl}$ (solid brown) in the single-TLS device for lower "0" and upper "1" energy levels spaced by $\epsilon _{0}$ for the 5 time periods. The duration of each ac field period is 1 ps. In the plot, the shown periods are separated by time intervals $\Delta T = 2 \times 10^{-10}$~s. The respective decay time of the single TLS device is estimated as $\tau _{\mathrm{dp}}=5 \times 10^{-9}$~s. The time decay of  $P_{0,1}$ peaks marks the loss of quantum coherence in the TLS device. The curves are computed using the Floquet-Markov (solid lines) and Lindblad (dashed blue and red lines) models at the temperature $T_{\max }=\allowbreak 580$ K assuming that $f=1$ THz. }}
\label{Fig_5}
\end{figure}

In calculations, we used the following parameters given in QuTiP units (QU). If we set, e.g., the clock frequency $f=1$ THz, then $\epsilon _{0}=\hbar \cdot 2\pi f = 4.1$~meV (we set it as 1 QU). Besides, $\omega = 6.3$~THz, $\delta =0.05$~QU $ = 0.18$~meV, and $A = 1.8$~meV. One ac field period is $T=1/f=10^{-12} $~s. The initial state for a single TLS setup is taken as $\psi _{0}=\left\vert \downarrow \right\rangle $, and we used the maximum temperature $T_{\max }= 580$~K.

As a measure of quantum coherence, we consider the time-dependent Rabi flops in plots of occupation probability versus time. In Fig.~\ref{Fig_3}a we plot the calculation results for the single period time dependence of occupation probabilities for lower ($P_{0}^{\rm Fl} $, curve~1) and upper ($P_{1}^{\rm Fl}$, curve~2) of energy levels spaced by $\epsilon _{0}=1$~QU. In Fig.~\ref{Fig_3} and below the time is in 1/QU units. Curves 1 and 2 are obtained using the Floquet-Markov formalism in the single TLS device. One can compare curves 1, and 2 with the dashed curves 3, and 4 illustrating the occupation probabilities $P_{0,1}^{\rm L}$ (curves ) obtained by solving the Lindblad equation. 

\begin{figure}
\includegraphics[width=125 mm]{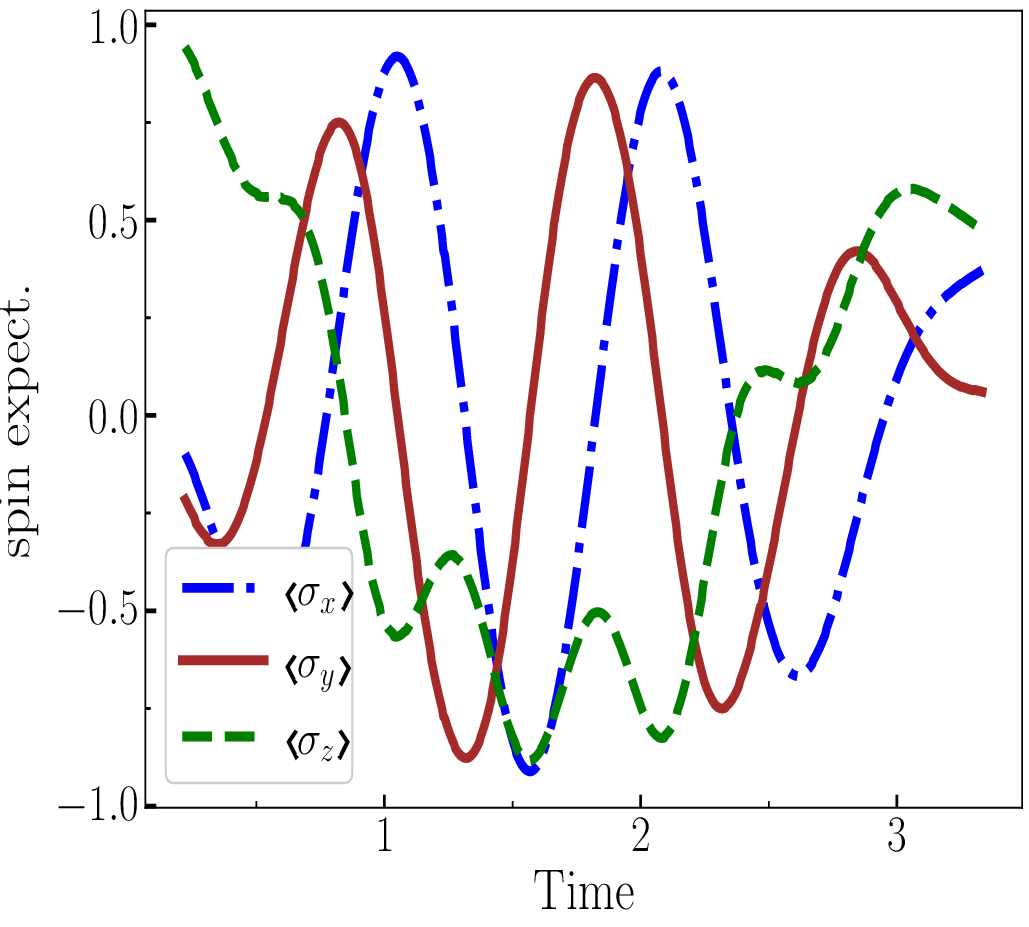} 
\caption{ { Expectation values $\left\langle \sigma _{x,y,z}\left( t\right)
\right\rangle $ for a single period of time. }}
\label{Fig_6}
\end{figure}

The other important characteristics are the expectation values of the spin operator $\sigma_j$ ($j = x ,y, z$) illustrating the spin dynamics in the single-TLS device depicted in Fig.~\ref{Fig_1}. In Figs.~\ref{Fig_3}b-d we also present the parametric plots of expectation values $\left\langle \sigma _{y}\right\rangle $ vs $\left\langle \sigma_{x}\right\rangle $, $\left\langle \sigma _{z}\right\rangle $ vs $\left\langle \sigma _{x}\right\rangle $, and also the one-time period dependence of $\left\langle \sigma _{x}\left( t\right) \right\rangle $.

\begin{figure}
\includegraphics[width=125 mm]{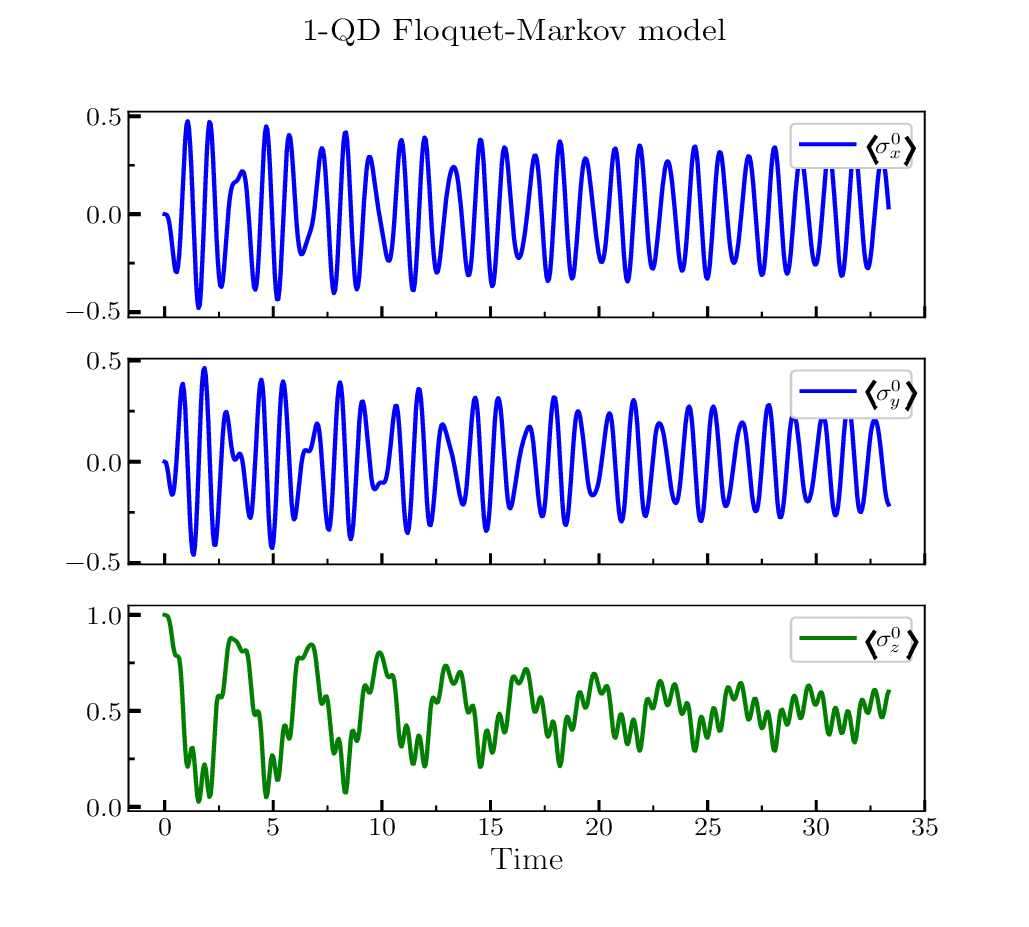} 
\caption{{Expectation values  $\left\langle \sigma _{x}^{(0)}\left( t\right) \right\rangle $ (upper pane), $\left\langle \sigma _{y}^{(0)}\left( t\right) \right\rangle $ (central pane), and $\left\langle \sigma _{z}^{(0)}\left( t\right) \right\rangle $ (lower pane). In this plot, to illustrate the spin dynamics on the short scale of just a few ac field periods, we use the much-increased noise intensity.
}}
\label{Fig_7}
\end{figure}

Further details of the spin dynamics are illustrated in Fig.~\ref{Fig_4} where in pane (a) we show the parametric plot of expectation values $\left\langle \sigma _{y}\right\rangle $ vs the occupation probability of lower level $P_{0}^{\rm Fl}$. The dependence $\left\langle \sigma _{y}\right\rangle $ vs $P_{0}^{\rm Fl}$ is complemented by the one-time period plot $\left\langle \sigma _{y} \left( t\right) \right\rangle $ presented in Fig.~\ref{Fig_4}b. Similar parametric plot $\left\langle \sigma _{x}\right\rangle $ vs $P_{0}^{\rm Fl}$ is shown in Fig.~\ref{Fig_4}c. The temperature effect of noise is evident in Fig.~\ref{Fig_4}d where we show the single period of the occupation probability of lower level $P_{0}^{\rm Fl}$ for three temperatures. One can see that the curves 1, 2, and 3 in Fig.~\ref{Fig_4}d, which correspond to temperatures $T = 33$~K (curve 1), $T = 280$~K (curve 2), and $T = 570$~K (curve 3) deviate from each other at the end of the period. Although the mentioned deviation is relatively low, it might become much stronger when certain conditions are met as we discuss below.

The next step is to compute the single-TLS characteristics on the long-time scale. We use the computed $\Phi _{\alpha }\left( t\in \lbrack 0,T]\right) $ to directly evaluate $\Phi _{\alpha }\left( t\right) $, $\Psi _{\alpha }\left( t\right) $, and $\Psi \left( t\right) $. This allows finding the occupation probabilities of interest. The obtained respective occupation probabilities $P_{0}$ (solid green) and $P_{1}$ (solid brown) for lower "0" and upper "1" energy levels spaced by $\epsilon _{0}$ are depicted in Fig.~\ref{Fig_5}. The time dependence $P_{0,1}(t)$ on the long scale is fairly monotonous and pieces of $P_{0,1}$ in adjacent periods are very similar to each other, which creates a presentation problem. 
\begin{figure}
\includegraphics[width=125 mm]{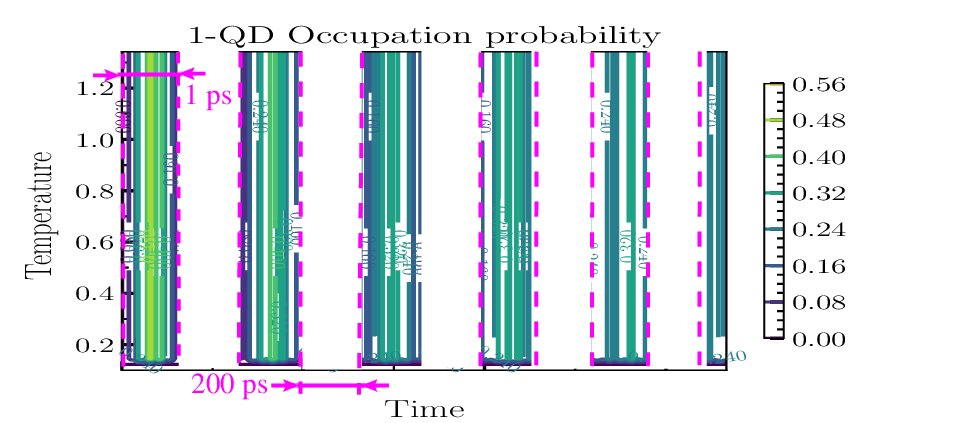} 
\caption{ {The contour plot illustrating the temperature dependence of the occupation probability $P_{0}^{\rm Fl}$ in the single-TLS device exposed to the external ac field. The temperature alters between $0$ and $580$ K assuming that $f=1$~THz. }}
\label{Fig_8}
\end{figure}
To see the temporal progress of $P_{0,1}(t)$, for the sake of better representation of the obtained results we show in Fig.~\ref{Fig_5} only 5 selected periods. We just skip the other periods in between. The duration of each period in Fig.~\ref{Fig_5} is 1~ps. In the plot, the shown periods are separated by rather large time intervals $\Delta T = 2 \times 10^{-10}$~s. The respective decay time $\tau _{\mathrm{dp}}$ of the single-TLS device is estimated as $\tau _{\mathrm{dp}} \simeq 5 \times 10^{-9}$~s. The curves are computed using the Floquet-Markov (solid lines) and Lindblad (dashed blue and red lines) models at the temperature $T_{\max }=\allowbreak 580$~K assuming that $f=1$~THz. 

We complement the single-TLS results with the expectation values $\left\langle \sigma _{x,y,z}\left( t\right) \right\rangle $ for a single period as shown in Fig.~\ref{Fig_6}.

In Fig.~\ref{Fig_7} we present the expectation values  $\left\langle \sigma _{x}^{(0)}\left( t\right) \right\rangle $ (upper pane), $\left\langle \sigma _{y}^{(0)}\left( t\right) \right\rangle $ (central pane), and $\left\langle \sigma _{z}^{(0)}\left( t\right) \right\rangle $ (lower pane) for five periods of time.

The temperature effect of noise is illustrated in the contour plot in Fig.~\ref{Fig_8}. To see the progress of the occupation probability $P_{0}^{\rm Fl}(t,T)$ on the long time scale we use the same trick as in Fig.~\ref{Fig_5}. Namely, we plot only selected five periods of $P_{0}^{\rm Fl}(t,T)$ separated from each other by 200 ps. One can see that $P_{0}^{\rm Fl}(t,T)$ considerably wanes out on the time scale $\sim 5$~ns, which is associated with the value of decay time $\tau_{\rm dc}$.

\subsection{The three-dot cluster driven by the ac field}\label{three-dot}

\begin{figure}
\includegraphics[width=125 mm]{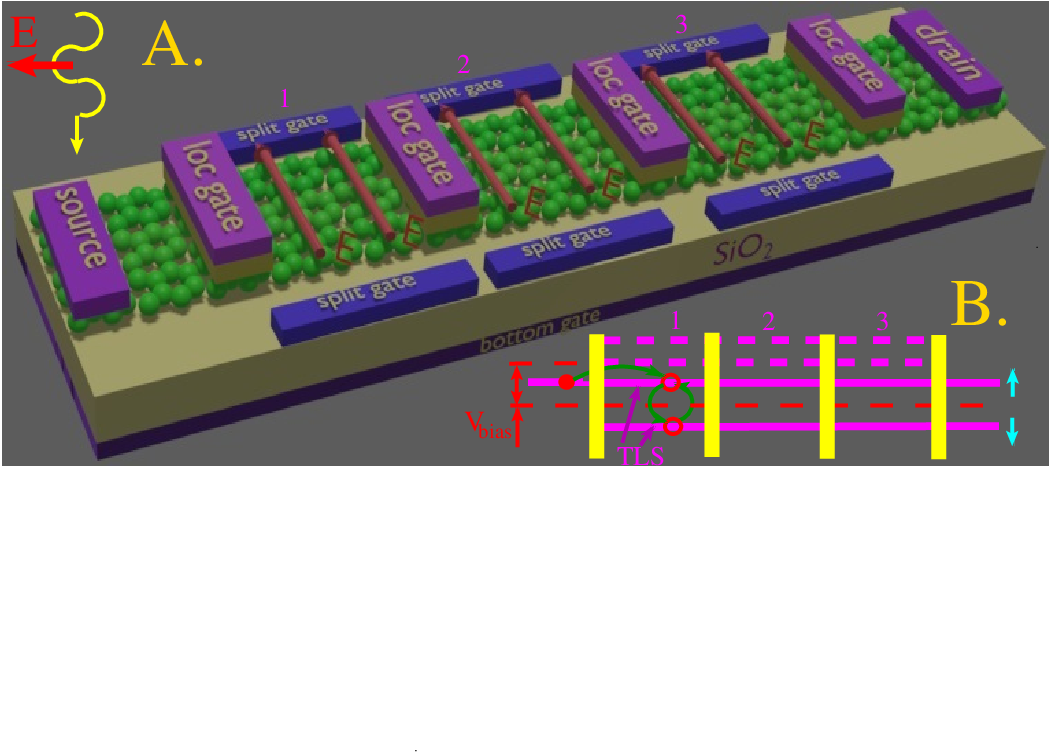} 
\caption{ { A.~Three quantum dots (marked as 1, 2, and 3) are formed on a graphene stripe with zigzag atomic edges (ZZ-stripe). The dots are separated from each other by the local gate electrodes, which control the inter-dot coupling strength by tuning the heights of the chiral barriers. The bias voltage $V_{\rm bias} $ applied to the source and drain electrodes induces electron interlevel transitions and causes interdot hoppings (see green arrows in the energy diagram B)  \cite{Shafr-Floq}. The split gates create a transversal electric field ${\bf E}$ causing the Stark splitting $\Delta $ of the zero-energy level into the two TLS levels marked by cian "$\uparrow $" and "$\downarrow $".  B.~The energy diagram illustrates the ac field-induced Floquet processes as discussed in SI,  Sec.~2. The dashed lines indicate the Floquet pseudoenergies. The bottom gate controls the mean value of the electrochemical potential $\mu $ in the ZZ-stripe.
}}
\label{Fig_9}
\end{figure}

Let's consider a numerical model of three coupled TLS depicted driven by the ac field as shown in Fig.~\ref{Fig_9}. 
The ac field not only drives the three-TLS system but also changes the effect of dissipation that becomes time-dependent. The stationary states of the complex quantum system have been considered in Refs.~\cite{Shafr-AQT-2020,Shafr-PRB-2023}. Earlier we found~\cite{Shafr-AQT-2020,Shafr-PRB-2023} that in the multi-TLS device, the energy spectrum experiences the intrinsic spectral narrowing (ISN) resulting in considerable shrinking of the quantized energy level width by 4-6 orders of magnitude. The ISN phenomenon leads to the reduction of the adverse noise influence by a few orders of magnitude causing the respective prolongation and resilience of the quantum coherence. Besides, a large spectral narrowing was obtained in the experimental work~\cite{Asada-2021} using the conventional approach. Below we focus on the dynamics of the three-TLS cluster: how the state of the system evolves with time. In unitary evolution, the state of the system remains normalized. In this non-unitary dissipative system, the state of the system does not remain normalized. While the evolution of the state vector in a closed quantum system is deterministic, the three-TLS open quantum systems are stochastic. The effect of an environment on the three-TLS system is to induce stochastic transitions between energy levels and between the dots inside the cluster, and to introduce uncertainty in the phase difference between states of the system. The state of the open quantum system is described in terms of ensemble-averaged states using the density matrix formalism. A density matrix $\rho (t)$ describes a probability distribution of quantum states $\left\vert \psi_n \right\rangle $, in a matrix representation $\rho =\sum_n p_n \left\vert \psi_n \right\rangle  \left \langle \psi_n \right\vert $, where $ p_n$ is the classical probability that the system is in the quantum state $\left\vert \psi_n \right\rangle $. In the remaining portions of this section, we focus on the time evolution of a density matrix $\rho (t)$.

\begin{figure}
\includegraphics[width=125 mm]{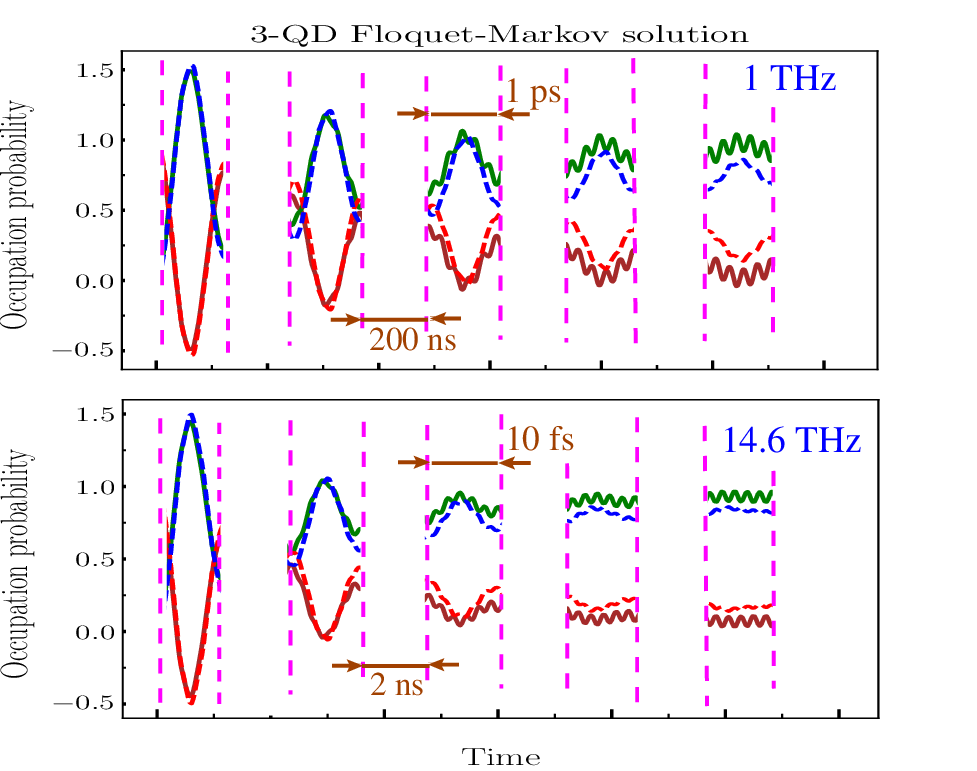} 
\caption{ { The time-dependent occupation probabilities $P_{0} $ (lower state) and $P_{1} $ (upper state) for the quantum dot cluster comprising three coupled TLS exposed to the external ac field. \emph{Upper pane}: The results obtained in the Floquet-Markov model (solid green and dark red curves ) for the triple TLS during 5 time periods at the temperature $580$~K assuming that $f=1$~THz. The 1~ps periods are separated by 200~ns intervals. We also show the results obtained by solving the Lindblad equation (blue and red dashed curves). The decay time is evaluated as $\tau_{\rm dc} \simeq 5 \times 10^{-6}$~s. The long $\tau_{\rm dc} $ at $580$~K occurs owing the relatively weak temperature dependence of $S(\omega, T)$ at  $f=1$~THz (see plots in Fig.~\ref{Fig_2}).  \emph{Lower pane}: Similar results as above but for the ac field with $f=14.6$~THz. As follows from Fig.~\ref{Fig_2}, the temperature dependence of the noise spectral density $S(\omega, T)$  is much stronger at  $f=14.6$~THz than it is at $f=1$~THz. That is why $\tau_{\rm dc} $ shortens down to $\tau_{\rm dc} \simeq 10$~ns when $f=14.6$~THz and $T = 580$~K. Similar shortening of $\tau_{\rm dc} $ occurs also for other ac frequencies as soon as $f$ coincides with the peak in $S(\omega, T)$ when the temperature dependence of the noise intensity becomes strong.
 }}
\label{Fig_10}
\end{figure}

To describe the states of multipartite quantum systems such as three coupled TLS we need to expand the Hilbert space by taking the tensor product of the state vectors for each of the system components. Similarly, the operators acting on the state vectors in the combined Hilbert space (describing the coupled system) are formed by taking the tensor product of the individual operators.

We also perform the partial trace representing an operation that reduces the dimension of a Hilbert space by eliminating some degrees of freedom by averaging (tracing). The partial trace operation acts as the converse of the tensor product and is used to focus only on the TLS part of the coupled quantum system while disregarding the details of the external environment. Practically, for the open quantum systems, this involves tracing over the environment leaving only the system of interest. 

We assume that in the three-TLS cluster, the participating quantum dots have equal energy splitting $\epsilon _{0}$ and that the adjacent TLS interact with each other through a $\sigma _{x} \otimes \sigma _{x}$ with strength $g$ (in units quoted below as QU where the bare TLS energy splitting is unity). 
The Hamiltonian describing the three coupled TLS is
\begin{equation}
{\cal H}=-\frac{\varepsilon_0}{2} \cdot (\sigma _{z}\otimes \hat{1}\otimes \hat{1}+\hat{1}\otimes \sigma
_{z}\otimes \hat{1}+\hat{1}\otimes \hat{1}\otimes \sigma _{z})  -\frac{\delta}{2} \cdot
\sigma _{x}\otimes \sigma _{x}\otimes \hat{1}+g\cdot \hat{1}\otimes
\sigma _{x}\otimes \sigma _{x}
\end{equation}%
where $\delta $ is the interlevel coupling energy. Interaction with the ac field with strength $A$ and angular frequency $\omega $ is given by
\begin{equation}
{\cal H}_{ac}=\frac{A}{2}\left( \sigma _{z}\otimes \hat{1}\otimes \hat{1}+2\cdot 
\hat{1}\otimes \sigma _{z}\otimes \hat{1}+\hat{1}\otimes \hat{1}\otimes
\sigma _{z}\right) \sin \left( \omega t\right) .
\end{equation}%
We start evolution with one of the TLS in its excited state%
\begin{equation}
\psi _{0}=\left\vert \downarrow \right\rangle \otimes \left\vert \uparrow
\right\rangle \otimes \left\vert \downarrow \right\rangle
\end{equation}%

To present the calculation results for the three-TLS cluster on the long-time scale we use the same trick as for the single-TLS device above. First, we compute $\Phi _{\alpha }\left( t\in \lbrack 0,T]\right) $ and use it to directly evaluate $\Phi _{\alpha }\left( t\right) $, $\Psi _{\alpha }\left( t\right) $, and $\Psi \left( t\right) $. Then we immediately compute the occupation probabilities $P_{0}$ and $P_{1}$ for lower "0" and upper "1" energy levels spaced by $\epsilon _{0}$. Due to the intrinsic narrowing phenomenon~\cite{Shafr-AQT-2020,Shafr-PRB-2023}, the time dependence $P_{0,1}(t)$ on the long scale is even much more monotonous than for the single-TLS device (see Sec.~\ref{single-TLS}). Because the pieces of $P_{0,1}$ in adjacent periods are very similar to each other, this creates a presentation problem because it's hard to distinguish them from each other. To resolve such a problem and illustrate the temporal progress of $P_{0,1}(t)$, we use the representation of obtained results to show in Fig.~\ref{Fig_10} only 5 selected periods. We just skip the other periods in between. Again, the duration of each period in the upper pane of Fig.~\ref{Fig_10} for the ac field frequency $f=1$~THz is 1~ps. But in this plot for $f=1$~THz and the period duration 1~ps, we show only periods separated by large time intervals $\Delta T = 2 \times 10^{-7}$~s. The periods in between are just not shown. We estimated the respective decay time $\tau _{\mathrm{dp}}$ of the three-TLS device as $\tau _{\mathrm{dp}} \simeq 5 \times 10^{-6}$~s. The $P_{0,1}(t)$-curves are obtained using the Floquet-Markov (solid lines) and Lindblad (dashed blue and red lines) equations at the temperature $T_{\max }=\allowbreak 580$~K.

\begin{figure}
\includegraphics[width=125 mm]{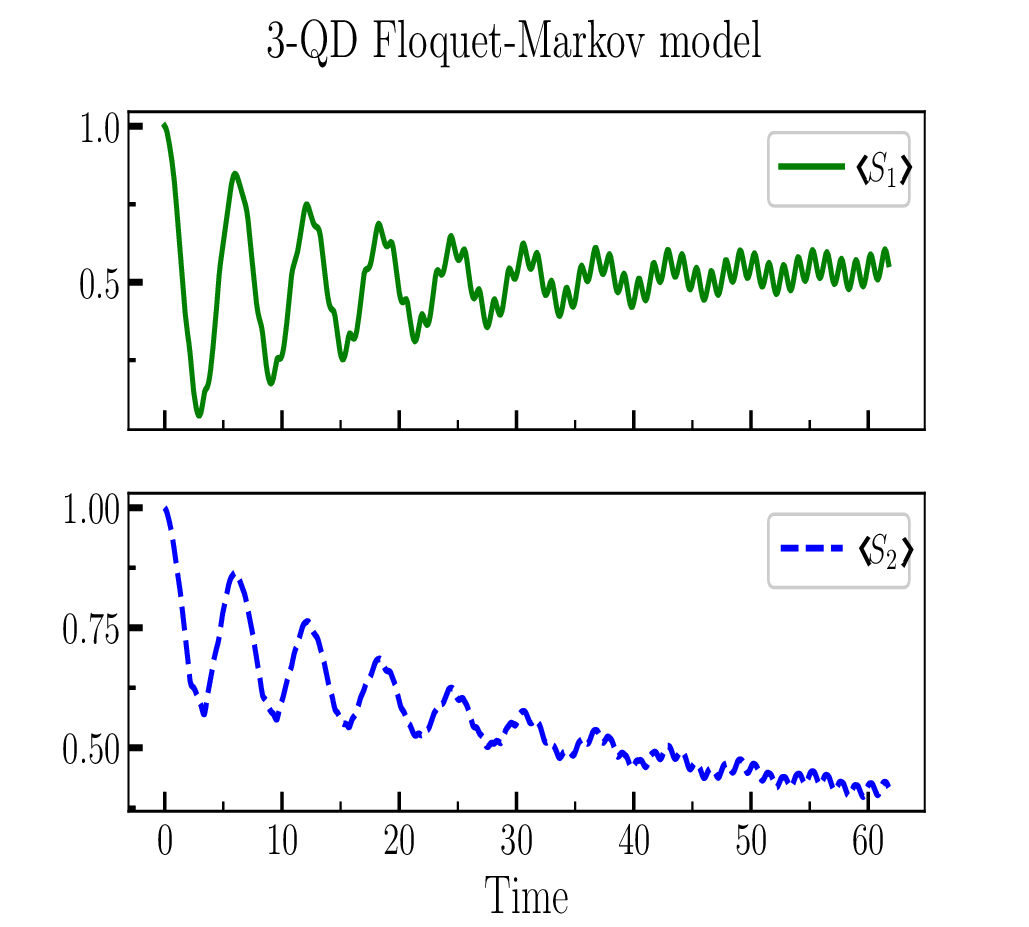} 
\caption{ {The time dependence of expectation values of the total spin components $\left\langle S_1 \left( t\right) \right\rangle $ and  $\left\langle S_2 \left( t\right) \right\rangle $ for the three-TLS device exposed to the external ac field. As in Fig.~\ref{Fig_7}, in this plot, we use the much-increased noise intensity, to illustrate the spin dynamics on the scale of just a few ac field periods.
}}
\label{Fig_11}
\end{figure}

In the same Fig.~\ref{Fig_10} we compare the data for the ac field frequency $f=1$~THz and the period duration 1~ps (shown in the upper pane) with the other data obtained for $f=14.6$~THz and the period duration 10~fs (see the lower pane). According to Fig.~\ref{Fig_2}, the temperature dependence of the noise spectral density $S(\omega, T)$  is much stronger at  $f=14.6$~THz than it is at $f=1$~THz. The much-increased noise intensity at  $f=14.6$~THz serves as the reason why $\tau_{\rm dc} $ shortens down to $\tau_{\rm dc} \simeq 10$~ns when $f=14.6$~THz. Analogous shortening of $\tau_{\rm dc} $ happens also for other ac frequencies as soon as $f$ coincides with the peak in $S(\omega, T)$ when the temperature dependence of the noise intensity becomes much stronger and the noise increases.

\begin{figure}
\includegraphics[width=125 mm]{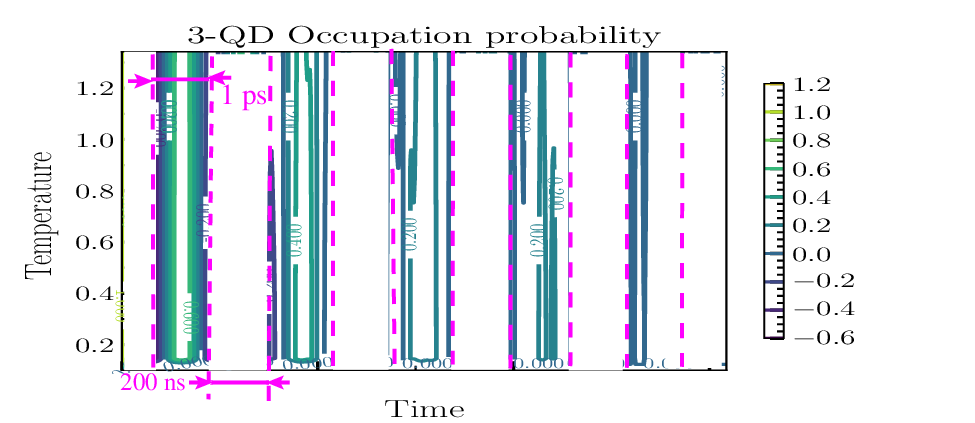} 
\caption{ {The contour plot illustrating the temperature dependence of the occupation probability $P_{0}^{\rm Fl}$ in the three-TLS device exposed to the external ac field. The temperature changes between $0$ K and $580$ K assuming that $f=1$~THz.
}}
\label{Fig_12}
\end{figure}

In Fig.~\ref{Fig_11} we show the time dependence of the expectation values of the total spin components $\left\langle S_1 \left( t\right) \right\rangle $ and  $\left\langle S_2 \left( t\right) \right\rangle $ for the three-TLS system. In particular, they characterize the time-dependent dissipation in the three-TLS cluster exposed to the external ac field. In this plot, to illustrate the spin dynamics on the scale of a few ac field periods, we use the much-increased noise intensity.

Fig.~\ref{Fig_12} illustrates the temperature and temporal dependence of the occupation probability $P_{0}^{\rm Fl}$ in the quantum dot cluster comprising three coupled TLS exposed to the external ac field.  The contour plot in Fig.~\ref{Fig_12} complements the data shown in Fig.~\ref{Fig_10} by detailing the temperature dependence of  $P_{0}^{\rm Fl}$. The temperature changes between $0$ K and $580$ K assuming that $f=1$~THz. One may notice that the temperature effect of noise is stronger for the three-TLS device than it was for the former one-TLS setup. Remarkably, $P_{0}^{\rm Fl}(t,T)$ wanes out considerably on the longer time scale $\sim 5$~$\mu $s, which is related to the longer decay time $\tau_{\rm dc}$ in the last case.

\begin{figure}
\includegraphics[width=125 mm]{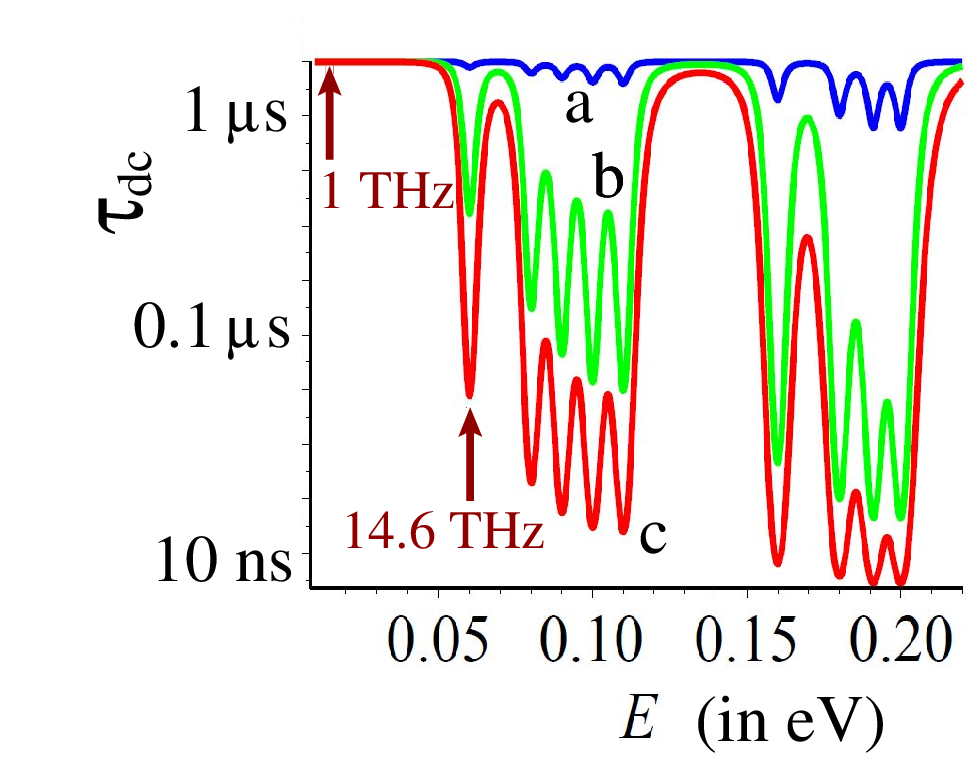} 
\caption{ {The energy dependence of decay time $\tau_{\rm dc} (E)$ for the three-TLS device at temperatures $T = 4.7$~K (blue curve a), $285$~K (green curve b), and $580$~K (red curve c). At $T = 580$~K (corresponds to $E = 0.2$~eV), the decay time significantly shortens down from $\tau_{\rm dc} (E) \simeq 5$~$\mu$s at $f = 1$~THz to just $\tau_{\rm dc} (E) \simeq 10$~ns when $f = 50$~THz.
}}
\label{Fig_13}
\end{figure}

In Fig.~\ref{Fig_13} we present the energy dependence of decay time $\tau_{\rm dc} (E)$ for temperatures $T = 4.7$~K (blue curve a), $285$~K (green curve b), and $580$~K (red curve c). One can see that $\tau_{\rm dc} (E)$ varies versus the temperature and energy. For flat regions of energy dependence, $\tau_{\rm dc} (E)$ behavior with the temperature is rather weak. The temperature dependence of the decay time becomes very significant when the energy $E$ coincides with peaks in the electron-phonon spectral density (compare with Fig.~\ref{Fig_2}). Respectively, in terms of the ac field frequency $f$, the decay time significantly shortens down from $\tau_{\rm dc} (E) \simeq 5$~$\mu$s at $f = 1$~THz (corresponds to $E = 4.1$~meV) to just $\tau_{\rm dc} (E) \simeq 10$~ns when $f = 50$~THz (i.e, $E = 0.2$~eV)  and $T = 580$~K.

\section{Discussion}\label{discuss}

The above-reported numerical experimental results concerning temperature limits of the transitional quantum dynamics in qubit clusters exposed to the ac field are summarized as follows.

The different sources of noise include Johnson-Nyquist (JN), photon, amplifier, load, and excessive noise. The JN noise is generated by the thermal agitation of the charge carriers (i.e.,  electrons) at equilibrium inside an electrical conductor with resistance $R$, which happens regardless of any applied voltage. Photon noise originates from the randomness in the signal associated with photons arriving at the TLS device. We neglected the photon noise by considering the purely harmonic signal with no randomness. The excess electrical noise involves two types, flicker (\textquotedblleft $1/f$\textquotedblright ) noise and contact shot noise, which may become significant at low frequencies. For the sake of efficiency of numeric calculations, all the noises were mimicked by a simple analytical expression.

Reduction of the external noise emerging from the external electrodes and dielectric substrate can be accomplished by using a bi-metal multilayer~\cite{Mayle-JP-2015}, which filters the heat flow by separating the electron and phonon components one from another. The multilayer structure minimizes the phonon component of the heat flow while retaining the electronic part. This improves the overall performance of the TLS cluster due to minimizing the energy dissipation. 

The temperature dependence of noise vs THz frequency is examined in this work using a model expression (\ref{Snoise}) mimicking the phonon spectrum of the graphene stripe. One appropriately selects parameters of the electron-phonon interaction, positions, and widths of phonon branches in  Eq.~(\ref{Snoise}), and also adds the term $S_{\rm other}\left( \omega, T \right) $ accounting for respective contributions from other noise sources such as Johnson-Nyquist (JN), photon, amplifier, load, and excessive noise. As is evident from Fig.~\ref{Fig_2}, the resulting noise spectral density $S\left( \omega, T \right) $ is frequency- and temperature-dependent. According the obtained resuls, presented in Figs.~\ref{Fig_3}-\ref{Fig_12}, in the flat portions of $S\left( \omega, T \right) $ the quantum coherence only weakly depends on the temperature. On the contrary, the temperature dependence of quantum coherence effects becomes considerable for the ac field frequencies coinciding with the peaks in $S\left( \omega, T \right) $. That conclusion is illustrated in Fig.~\ref{Fig_13}, where we presented the temperature effect on quantum coherence. The data shown in Fig.~\ref{Fig_13} are obtained from the sequence of contour plots $P_0(\omega,T,t) $ illustrating the suppression of quantum coherence by noise. From Figs.~\ref{Fig_2} and \ref{Fig_13} one can observe that noise intensity rather weakly depends on the temperature at flat regions of $S\left( \omega, T \right) $. On the contrary, at the phonon frequencies $\omega \sim \Omega_n $ ($\Omega_n $ is the frequency of the $n$-th phonon spectrum branch), the noise intensity increases about by the factor of five when the temperature rises from $T = 0$~K to $T = 580$~K. Thus, if the external ac field frequency coincides with $\Omega_n $, the adverse influence of noise on the quantum coherence considerably intensifies. Respectively, at $\omega \simeq \Omega_n $, the decay time $\tau_{\rm dc} $ is shortened.

The role of spectral narrowing is that either the conventional~\cite{Asada-2021} or intrinsic spectral narrowing~\cite{Shafr-AQT-2020,Shafr-PRB-2023} of the quantized energy levels help to reduce the effect of decoherence by several orders of magnitude. This causes a considerable shrinking of the quantized energy level width by 4-6 orders of magnitude. The mentioned phenomenon ultimately leads to the reduction of the adverse noise influence by a few orders of magnitude causing the respective prolongation and resilience of the quantum coherence in the multi-TLS devices.

Using a simple model we found conditions when maintaining the quantum coherence is possible even at high temperatures exceeding the room temperature. A feasible strategy is to diminish the impact of noise on the quantum coherence that is by using a combined approach involving (i) the bi-metal multilayers protecting the TLS device from the external noise~\cite{Mayle-JP-2015}, (ii) the spectral narrowing effect~\cite{Asada-2021,Shafr-AQT-2020,Shafr-PRB-2023}, and (iii) avoiding the operation at frequencies coinciding with peaks in $S_{\rm other}\left( \omega, T \right) $.

\section{Conclusions}\label{conclude}

The performed numerical experiments of the transitional quantum dynamics in the all-electrically controlled single- and multi-TLS devices exposed to the electromagnetic field indicate a dramatic interplay affecting the temperature limits of the noise effect on the intrinsic coherence. The results obtained in the Floquet-Markov model and Lindblad equation allow quantitative conclusions on how the external ac field changes the coherent transport in QC and also how it depends on the spectral density of noise. We find the conditions when the quantum coherence in QC is prolonged by several orders of magnitude even above the room temperature which can serve to improve the performance of various multi-TLS devices. As an example, we considered the three-TLS device and found that the temperature dependence of the decay time significantly shortens down from $\tau_{\rm dc} (E) \simeq 5$~$\mu$s for the ac field frequency $f = 1$~THz to just $\tau_{\rm dc} (E) \simeq 10$~ns when $f = 50$~THz  and $T = 580$~K. The results can be utilized when developing novel multi-TLS devices used in various areas of science and technology such as  terahertz sensors, analyzers, lasers, and quantum computing circuits.


\renewcommand{\theequation}{\arabic{equation}-Apx}

\section{Appendix}\label{appendix}

\subsection{The electron-phonon collisions in graphene nanoribbons}\label{e-ph}

The electron-phonon collisions generate temperature-dependent noise, thus adversely affecting the quantum coherence. We evaluate the golden-rule decay rate of the TLS similarly to the case of as described below. The source of incoherence originating from the inelastic electron-phonon scattering is characterized by the rate $\gamma _{\mathrm{ep}} $, which is explained below. The energy dissipation in the TLS circuit depends on the inelastic scattering rate $\gamma _{\mathrm{ep}} $, which determines the coherence time $\tau _{\mathrm{c}}$ of the TLS. Typical value of the electron-phonon coupling constant in graphene $\lambda _{\mathrm{ep}}^{\mathrm{G}}=0.1-0.35$~\cite{Benedek} is rather low as compared to typical superconducting metals, where $\lambda _{\mathrm{ep}}=0.4-1.3$~ \cite{PAllen}. Therefore, we assume that the weak electron-phonon coupling in graphene is described in the linear response approximation. For the graphene TLS, the most important processes of the electron-phonon scattering involve optical phonons with finite energy $\hbar \omega _{\mathrm{opt}}$ but zeroth momentum $q$. 

Besides, the phonon spectrum of graphene stripes depends on the shape of atomic edges and the stripe width~\cite{Karamita,YWang}. E.g., for the graphene stripe with zigzag-shaped atomic edges by width $W=1$~nm, the number of optical phonon branches per energy interval 0-50~meV is four~\cite{Karamita}, which gives the energy spacing between the optical phonon branches as $\Delta \omega _{\mathrm{opt}}\sim 12$~meV. For the wider graphene stripe by width $W=3$~nm one gets 9 optical phonon branches spaced by 5.5~meV. The electron level spacing $\Delta _{n}$ varies and is controlled by applying
the electric voltage to either the split gate or local gate as shown in Fig.~1 of main text (see details in Ref.~\cite{Shafr-PRB-2023}). Technically, the noise spectral-density function of the environment is implemented in QuTiP as a Python callback function that is passed to the solver. 

A simple recipe for either eliminating or significantly reducing the energy dissipation due to the electron-phonon scattering is as follows: (a) use just these two energy levels, whose positions and the inter-level spacing are electrically controlled by the gate voltage, (c) select the energy levels whose energies $E_{1,2}$ after adjustment by the phonon energy $\hbar \omega _{A_{1}^{\prime }}$ do not coincide with $E_{3,4}+\hbar \omega_{A_{1}^{\prime }}$, i.e., 
\begin{equation}
E_{3,4}+\hbar \omega _{A_{1}^{\prime }}\neq E_{1,2}.  \label{match}
\end{equation}%
The above condition~(\ref{match}) is broken when 
\begin{equation}
E_{3,4}+\hbar \omega _{A_{1}^{\prime }}=E_{1,2}\mbox{.}  \label{mismatch}
\end{equation}%
One can see that the coherence is immediately destroyed due to the electron-phonon scattering. The latter condition~(\ref{mismatch}) can be exploited to protect a TLS against external influence or to segregate adjacent TLS. From the above, it is clear that the electron-phonon scattering in TLS occurs only when the positions of electron energy levels match the
condition~(\ref{mismatch}). When the condition~(\ref{mismatch}) is not satisfied, the electron-phonon interaction is diminished or even vanishes.

The effective width $\Gamma _{\mathrm{FF}}$ of electron energy level related to the decay rate $\gamma_{\rm dc}=\Gamma_1 + \Gamma_2$ of TLS device, can be significantly diminished using the effect of intrinsic spectral narrowing~\cite{Shafr-AQT-2020,Shafr-PRB-2023}. The decay time then is $\tau_{\rm dc} = \hbar/\gamma_{\rm dc}$. In graphene TLS, inside the four-dot qubit cluster, $\Gamma _{\mathrm{FF}}$ may be reduced by several orders of magnitude, becoming $\Gamma _{\mathrm{FF}}\simeq 1.3\times 10^{-5}$~$\mu $eV \cite{Shafr-AQT-2020,Shafr-PRB-2023}. Since $\Gamma _{\mathrm{FF}} << \min {\{\Delta \omega _{\mathrm{opt}}, \Delta _{n}\}}$, one can conform/violate the above condition~(\ref{match}) [or otherwise~(\ref{mismatch})] by merely adjusting the electron energy level positions $E_{n,m} $ (where $n$ and $m$ are the respective level indices) either by changing the interdot barrier height or by adjusting the Stark splitting magnitude by applying appropriate gate and source-drain voltages. The phonon branch positions for each particular graphene quantum dot configuration can also be detected experimentally by measuring the differential conductance, whose anomalies at certain values of the bias voltage would indicate when the electron-bound state decays due to inelastic electron-phonon scattering.

The above recipes can be used to design the noise-proof multi-TLS devices, where one avoids the undesirable energies $E_{\mathrm{LO/TO}}=E_{3,4}-\hbar \omega _{\mathrm{LO/TO}}$ and $E_{A_{1}^{\prime }}=E_{3,4}-\hbar \omega_{A_{1}^{\prime }}$ as soon as the dissipation becomes too large in the multi-TLS devices. One accomplishes this by appropriately selecting the local gate voltage $V_{\lg }$ (see sketches in Figs.~1, 9 of main text) to ensure that the level splitting $\Delta _{n} $ is such that $E_{\mathrm{LO/TO}}$ and $E_{\mathrm{{A}_{1}^{\prime }}}$ don't coincide with any electron energy level. Otherwise, when there is a need to isolate the TLS, one sets the level to coincide with $E_{\mathrm{LO/TO}}$ and $E_{\mathrm{{A}_{1}^{\prime }}}$. The above means that for achieving the room-temperature functionality, one should use much smaller values of $\Delta \sim 30$ ~meV, which respectively corresponds to graphene stripe width considerably below $W=\hbar v_{\mathrm{F}}/(\pi \Delta_{n} )\sim 20$~nm.

The ultimate decay rate $\gamma_{\rm tot} $ of the TLS's quantum state is determined by the influence of external noise and additionally by the dissipative processes of inelastic scattering, $\gamma_{\rm tot} =\Gamma _{1}+\Gamma _{2}+\gamma _{\mathrm{ep}} $. Technically, the interaction of the noise field with electrons in the quantum dot is described similarly to the electron-phonon interaction~\cite{Zanker}. The relevant microscopic process is temperature-dependent because it represents an inelastic scattering involving changes in the electron energy and momentum. However, during the scattering in the low-dimensional device such as the quantum dot, the momentum and energy conservation laws impose additional constraints on the process probabilities, resulting in the eventual diminishing of $\gamma_{\rm tot} $ due to the following. (i) Conservation of the electron momentum $\mathbf{p}$ requires that its change $\delta \mathbf{p}=0$, (ii) the energy conservation requires that the energy of a noise quantum must match the level spacing $\Delta _{n}$. The electron chirality conservation in graphene TLS introduces additional selection rules since the electron momentum and energy change during its scattering in the K-point vicinity must oblige those energy and momentum conservation rules. Thus, the temperature dependence of the dissipation processes is reduced or even vanishes, protecting the graphene quantum dot TLS circuit against thermal fluctuations at elevated temperatures. The most important constraint in ZZ-TLS is the narrow width $\Gamma_{\rm FF} $ of the energy levels, i.e., $\Gamma_{\rm FF} <<\Delta_{n} $. This assumes that the energy dissipation does not occur unless the phonon line coincides with the bound state level.

To achieve the room temperature functionality of the TLS device, the
separation between two adjacent edge state levels must exceed $\Delta
_{n}=\hbar v_{F}/(\pi W)\approx 30$~meV provided the graphene stripe width $%
W=\hbar v_{F}/(\pi \Delta _{n})<20$~nm. The TLS interact with and thus
dissipate information into the noisy environment, introducing differences to
the ideal result because the interaction with the environment adds a
perturbation resulting in the TLS's dephasing and relaxation~\cite{Zanker}.
The uncertainty due to noise arises in addition to the temperature-dependent
inelastic scattering such as electron-phonon collisions. In the worst case
scenario, the total electron energy uncertainty is estimated as $\gamma_{\rm tot}
= \Gamma _{1}+\Gamma _{2}+\gamma _{\mathrm{ep}}\approx \left(
0.01-0.03\right) \Delta_{n} $. For the sake of the numeric calculation efficiency, instead of operating with the dephasing and relaxation time, we will introduce the common decay time $\tau_{\rm dc}$ which is associated with  $\hbar/\gamma_{\rm tot} $ (here we use units with $\hbar = 1$). The calculation details of the electron spectrum
are given in Refs.~\cite{Shafr-AQT-2020,Shafr-PRB-2023}.

Relevant inelastic scattering mechanisms in graphene involve the electrons scattering on acoustic phonons and optical phonons~\cite%
{Karamita,Nika,Ando-e-ph-scatter-2009,Das-Sarma-Mobility-Graphene-2008,nanomaterials-10-00039, TEbook,Munoz,Sanders,Savin}. In the graphene quantum dot clusters and arrays, this process can be readily avoided by applying appropriate local gate voltage and introducing the required mismatch. Below we disregard the electron-electron collisions on the timescale of decay since the electron density in the quantum dot systems of interest is relatively low.

For the electron-phonon scattering, the electron relaxation time $\tau_{\rm ep} $ depends on the electron density of states $N\left( \varepsilon _{k}\right) $ and temperature $T$~\cite%
{Karamita,Nika,Ando-e-ph-scatter-2009,Das-Sarma-Mobility-Graphene-2008,nanomaterials-10-00039, TEbook,Munoz,Sanders,Savin}. For the pristine graphene, one distinguishes several regimes~\cite%
{Sohier-thesis-2016} such as the Bloch-Gr\"{u}neisen (BG) regime,
equipartition (EP) regime, and high temperature (HT) regime. The BG regime takes place at $0$~K$<T\leq 0.15\times T_{\mathrm{BG}}$, $k_{B}T_{\mathrm{BG}} = 2\hbar k_{F}v_{TA/LA}$, where $v_{TA/LA}$ is the sound velocity of the TA/LA branches (typically, $v_{TA}=13.6$~km/s and $v_{LA}=21.4$~km/s). At the relevant temperatures, $k_{B}T$ is too small compared to the energy of optical phonons, thus their contribution is negligible, while the acoustic modes contribute since $k_{B}T$ is of the order of  the phonon energy $\hbar \omega _{q,TA/LA } $. Furthermore, the occupation of initial states $f\left( \varepsilon_{k}\right) $ and scattered states $f\left( \varepsilon _{k}\pm \hbar \omega _{q,TA/LA}\right) $ are significantly different. Here $ \varepsilon _{k}$  is the electron energy. In the EP regime at $0.15\times T_{\mathrm{BG}}\leq T\leq $ $\hbar \omega _{A_{1}^{\prime }}/k_{B}\approx 270$ K, optical phonons do not contribute into inelastic scattering but because $\hbar \omega _{q,TA/LA} << k_{B}T <<\varepsilon _{F}$, the scattering by acoustic phonons can be approximated as elastic. In the HT regime taking place at $T\geq 0.15\times \hbar \omega _{A_{1}^{\prime }}/k_{B}\approx 270$~K, the elastic approximation for acoustic phonons is still valid. Still, in the case of optical phonons, the three energy scales are comparable with each other. Hence, no suitable approximation can be made globally. Since the energy of optical $A_{1}^{\prime }$ phonons is lower than the LO/TO phonons and they couple stronger, the contribution of the former is higher than the latter.

Provided the optical phonons are hardly excited even at room temperature, the phonon emission process is dominant, and hence the scattering rate in GC is given by%
\begin{equation}
\gamma _{\mathrm{ep}}\left( \varepsilon \right) =2\pi ^{2}\lambda _{\mathrm{%
ep}}^{\mathrm{G}}\hbar ^{2}v^{2}N\left( \varepsilon -\hbar \omega \right) %
\mbox{,}  \label{gamma1}
\end{equation}%
where $N$ is normalized to $\Delta /(2\pi \hbar ^{2}v^{2})$ and%
\begin{equation}
\lambda _{\mathrm{ep}}^{\mathrm{G}}=\frac{36\sqrt{3}}{\pi }\frac{\hbar ^{2}}{%
2Ma^{2}}\frac{1}{\hbar \omega }\left( \frac{\beta }{2}\right) ^{2}    \mbox{.}
\label{lambda}
\end{equation}%
For zone-center phonons, $\omega _{\Gamma }=196$~meV and $\lambda _{\mathrm{ep}}^{\mathrm{G,\Gamma }}=2.9\times 10^{-3}(\beta _{\Gamma }/2)^{2}$, while $\omega _{K}=161.2$~meV and $\lambda _{\mathrm{ep}}^{\mathrm{G,K}} = 3.5\times 10^{-3}(\beta _{K}/2)^{2}$ for zone-boundary phonons, suggesting that zone-boundary phonons dominate over zone-center phonons. Thus, the phonon frequency is the unique parameter that determines the electron lifetime~\cite{Ando-e-ph-scatter-2009}. According to Ref.~\cite{Benedek}, the average over the available data gives for pristine graphene $\lambda _{\mathrm{ep}}^{\mathrm{G,Gr}}=0.22-1.1$, depending on the substrate. In pristine graphene, the electron-phonon scattering time is obtained at $T=300$~K as $\tau _{\mathrm{ep}}=\gamma _{\mathrm{ep}}^{-1}\simeq 10$~ps~\cite{Gunst}. In the graphene quantum dots, due to additional constraints on the permitted scattering processes in Eq.~(\ref{gamma1}) we use the effective values $\lambda _{\mathrm{ep}}^{\mathrm{G}}=0.1$ and $2\pi \hbar ^{2}v^{2}N\left( \varepsilon -\hbar \omega \right) \simeq 10^{-6}$~eV. In Eqs.~(\ref{gamma1})-(\ref{lambda}) we have generalized the results of Refs.~\cite{Ando-e-ph-scatter-2009,Sohier-thesis-2016} on the quantum dot geometry. Provided the optical phonons are hardly excited even at room temperature, the phonon emission process is dominant, and hence the scattering rate~\cite{Ando-e-ph-scatter-2009} is estimated as $\gamma _{\mathrm{ep}}\simeq \allowbreak 3\times 10^{-7}$~eV, which gives $\tau _{\mathrm{ep}}=\hbar /\gamma _{\mathrm{ep}}\simeq 2\times 10^{-9}$~s. The obtained coherence time $\tau _{\mathrm{c}}\simeq \tau _{\mathrm{ep}}=2$~ns is considerably prolonged due to the intrinsic spectral narrowing~\cite{Shafr-AQT-2020,Shafr-PRB-2023} up to $5\times 10^{-5}$~s in a multi-TLS cluster formed on stripes with zigzag atomic edges.

\subsection{Lindblad master equation}\label{Linbl}

When the ac field acts on the qubit cluster, it affects not only the electron states but also modulates the dissipation, which becomes time-dependent. Therefore, the conventional Lindblad master equation is not an adequate description of the system anymore. A more consistent approach should explicitly consider the time dependence of the driving field.  There are various approaches to solving the mentioned problem, but one general method \cite{Grifoni-1998,Shevchenko, Creffield-2003} is to derive the master equation in the Floquet basis.

The time-periodic alternating field is taken into account in Hamiltonian as
\begin{equation}
{\cal H}(t)={\cal H}_{0}+{\cal H}_{1}\sin (\omega t+\phi _{0})   \label{Hsh}
\end{equation}%
where ${\cal H}_{0}$ and ${\cal H}_{1}$ are two non-commuting time-independent terms, $\omega $ is the frequency of the alternating field, and $\varphi_{0}$ is an initial phase. An illustrative example is a two-level system (TLS) that interacts with an oscillating electric or magnetic field. In practice, this happens when driving transitions with a laser or microwave field. Below we consider the influence of an ac field on the transitional quantum dynamics of the TLS system in terms of Lindblad master equation and Floquet-Markov formalism. There is a difference in the evolution of the state vector in a closed quantum system on the one hand and in the open quantum system on the other hand: in the first case, it is deterministic, while it becomes stochastic in the second case. The external environment influences our system by
inducing stochastic transitions between energy levels, thus introducing uncertainty in the phase difference between states of the system. For the description of an open quantum system, one uses ensemble-averaged states in terms of the density matrix formalism. A probability distribution of quantum states using density matrix $\rho $ is $\left\vert \psi _{n}\right\rangle $, in a matrix representation $\rho =\sum_{n}p_{n}\left\vert \psi _{n}\right\rangle \left\langle \psi _{n}\right\vert $, where $p_{n}$ represents the classical probability that the system is in the quantum state of  $\left\vert \psi _{n}\right\rangle $. The above formalism allows us to describe the time evolution of a density matrix $\rho $.

The equations of motion for a system interacting with the environment are obtained by expanding the scope of the system to include the environment. Then the combination of such two systems becomes a unified and closed quantum system, whose evolution is described by the von Neumann equation%
\begin{equation}
\dot{\rho}_{\mathrm{tot}}\left( t\right) =-\frac{i}{\hbar }\left[ {\cal H}_{\mathrm{%
tot}},\rho _{\mathrm{tot}}\left( t\right) \right]  \label{vNeumann}
\end{equation}%
which is the equivalent of the Schr\"{o}dinger equation in the density matrix formalism. The total Hamiltonian includes the original system Hamiltonian ${\cal H}_{\mathrm{sys}}$, the Hamiltonian for the environment ${\cal H}_{\mathrm{env}}$, and a term representing the interaction between the system and its environment ${\cal H}_{\mathrm{int}}$%
\begin{equation}
{\cal H}_{\mathrm{tot}}={\cal H}_{\mathrm{sys}}+{\cal H}_{\mathrm{env}}+{\cal H}_{\mathrm{int}}.
\end{equation}%
Our interest is focused on the system dynamics, therefore now we can conduct a partial trace over the environmental degrees of freedom in Eq.~(\ref{vNeumann}) that yields a master equation for the motion of the original system density matrix. The most general trace-preserving and completely positive form of this evolution results in the Lindblad master equation for the reduced density matrix $\rho =\mathrm{Tr}_{\mathrm{env}}\left[ \rho _{%
\mathrm{tot}}\right] $%
\begin{equation}
\dot{\rho}\left( t\right) =-\frac{i}{\hbar }\left[ {\cal H},\rho \left( t\right) %
\right] +\sum_{n}\frac{1}{2}\left[ 2C_{n}\rho \left( t\right) C_{n}^{\dagger
}-\rho \left( t\right) C_{n}^{\dagger }C_{n}-C_{n}^{\dagger }C_{n}\rho
\left( t\right) \right]    \label{LindbladEq}
\end{equation}%
where $C_{n}=\sqrt{\gamma _{n}}A_{n}$ are collapse operators, and $A_{n}$ are the operators through which the environment couples to the system in ${\cal H}_{%
\mathrm{int}}$, and $\gamma _{n}$ are the corresponding rates. The derivation details of Eq.~(\ref{LindbladEq}) is given in various books \cite{Breuer,Haug,Nielsen}. 

If for an open system, the above conditions are satisfied, the Lindblad master equation~(\ref{LindbladEq}) gives an ensemble average of the system dynamics by describing the time evolution of the system density matrix. To make sure that the above approximations remain valid, one checks whether the decay rates $\gamma _{n}$ are smaller than the minimum energy splitting in the system Hamiltonian. A special caution must be paid to the systems strongly interacting with the environment and systems with degenerate or nearly degenerate energy levels.

The master equations represent the common approach for governing the non-unitary evolution of a quantum system, i.e., an evolution that includes incoherent processes such as relaxation and dephasing. In this work we utilize QuTiP solvers  \cite{QuTiP1,QuTiP2}, and the function qutip.mesolve is used for the evolution according to the Schr\"{o}dinger equation as well as the master equation, despite these two equations of motion strongly differ from each other. The \emph{qutip.mesolve} function automatically determines if it is sufficient to use the Schr\"{o}dinger equation (if no collapse operators were given) or if it has to use the master equation (if collapse operators were given). In many cases, calculating the time evolution according to the Schr\"{o}dinger equation is easier and much faster (for large systems) than using the master equation. Thus, if it becomes justified, the solver will automatically fall back on using the Schr\"{o}dinger equation.

The description of dissipation in the quantum system due to its interaction with an environment represents the new element in the master equation compared to the Schr\"{o}dinger equation. The respective interactions with the environment are defined using the operators describing the interaction of the system with the environment, which are complemented by rates that describe the strength of the processes.

\subsection{Floquet-Markov formalism}\label{sec-Fl-Mark}

In this Section, we briefly describe the Floquet-Markov formalism for solving time-dependent problems in QuTiP \cite{QuTiP1,QuTiP2}. The Schr\"{o}dinger equation with a time-dependent Hamiltonian ${\cal H}(t)$ is%
\begin{equation}
i\hbar \frac{\partial }{\partial t}\Psi \left( t\right) ={\cal H}\left( t\right)
\Psi \left( t\right)            \label{Sht}
\end{equation}%
where $\Psi \left( t\right) $ is the wave function solution and ${\cal H}(t)$ is given by Eq.~(\ref{Hsh}). We solve a problem with periodic time-dependence, i.e., when the Hamiltonian satisfies ${\cal H}\left( t\right) = {\cal H}\left( t+T\right) $ where $T$ is the period. According to the Floquet theorem, there exist solutions to~(\ref{Sht}) in the form
\begin{equation}
\Psi _{\alpha }\left( t\right) =\exp \left( -i\epsilon _{\alpha }t\right)
\Phi _{\alpha }\left( t\right)  \text{,}       \label{Psi}
\end{equation}%
where $\Psi _{\alpha }\left( t\right) $ are the Floquet states representing the set of wave function solutions to the time-dependent Schr\"{o}dinger equation, $\Phi _{\alpha }\left( t\right) =\Phi _{\alpha }\left( t+T\right) $ are the time-periodic Floquet modes, and $\epsilon _{\alpha }$ are the quasienergy levels. The quasienergy levels are constants in time, but only uniquely defined up to multiples of $2\pi /T$ (i.e., unique value in the interval $[0,2\pi /T]$). Provided the Floquet modes (for $t\in \lbrack 0,T]$) and the quasienergies for a particular ${\cal H}\left( t\right) $ are known, we can readily decompose any initial wavefunction $\Psi \left( t=0\right) $ in the Floquet states and then immediately find the solution for any $t$ as%
\begin{equation}
\Psi \left( t\right) =\sum_{\alpha }c_{\alpha }\Psi _{\alpha }\left(
t\right) =\sum_{\alpha }c_{\alpha }\exp \left( -i\epsilon _{\alpha }t\right)
\Phi _{\alpha }\left( t\right) \text{,}    \label{Fl-mod}
\end{equation}%
where the coefficients $c_{\alpha }$ are obtained from the initial
wavefunction $\Psi \left( 0\right) =\sum_{\alpha }c_{\alpha }\Psi _{\alpha }\left( 0\right) $. This approach helps finding $\Psi \left(t\right) $ for a given ${\cal H}\left( t\right) $ provided one can obtain the Floquet modes $\Phi _{\alpha }\left( t\right) $ and quasienergies $\epsilon _{\alpha }$ more easily than directly solving (\ref{Hsh}). By substituting (\ref{Psi}) into the Schr\"{o}dinger equation (\ref{Sht}) we arrive at an eigenvalue equation for the Floquet modes and quasienergies%
\begin{equation}
\rm{H}\left( t\right) \Phi _{\alpha }\left( t\right) =\epsilon _{\alpha
}\Phi _{\alpha }\left( t\right) \text{,}     \label{Eig}
\end{equation}%
where ${\rm H}\left( t\right) ={\cal H}\left( t\right) -i\hbar \partial _{t}$. The eigenvalue problem (\ref{Eig}) is solved either analytically or numerically. An alternative approach \cite{Creffield-2003} for numerically finding the Floquet states and quasienergies is used in QuTiP as follows. One introduces the propagator for the time-dependent Schr\"{o}dinger equation (\ref{Hsh}), which satisfies%
\begin{equation}
U\left( T+t,t\right) \Psi \left( t\right) =\Psi \left( t+T\right) \text{.}
\end{equation}%
Inserting the Floquet states from (\ref{Psi}) into the above expression gives%
\begin{equation}
U\left( T+t,t\right) \exp \left( -i\epsilon _{\alpha }t\right) \Phi _{\alpha
}\left( t\right) =\exp \left( -i\epsilon _{\alpha }\left( T+t\right) \right)
\Phi _{\alpha }\left( t+T\right) \text{,}
\end{equation}%
or using that $\Phi _{\alpha }\left( t+T\right) =\Phi _{\alpha }\left( t\right) $, one finds
\begin{equation}
U\left( T+t,t\right) \Phi _{\alpha }\left( t\right) =\exp \left( -i\epsilon
_{\alpha }T\right) \Phi _{\alpha }\left( t\right)
\end{equation}%
which shows that the Floquet modes are eigenstates of the one-period propagator. Therefore one finds the Floquet modes and quasienergies $\epsilon _{\alpha }=-\hbar \arg \left( \eta _{\alpha }\right) /T$ by numerically calculating $U\left( T+t,t\right) $ and diagonalizing it. This allows obtaining $\Phi _{\alpha }\left( 0\right) $ by calculating and diagonalizing $U\left( T,0\right) $.

For an arbitrary time moment $t$, the Floquet modes are found by propagating $\Phi _{\alpha }\left( 0\right) $ to $\Phi _{\alpha }\left( t\right) $ using the wave function propagator $U\left( t,0\right) \Psi _{\alpha }\left(0\right) $, which for the Floquet modes yields
\begin{equation}
U\left( t,0\right) \Phi _{\alpha }\left( 0\right) =\exp \left( -i\epsilon
_{\alpha }t\right) \Phi _{\alpha }\left( t\right)
\end{equation}%
so that $\Phi _{\alpha }\left( t\right) =\exp \left( i\epsilon _{\alpha }t\right) U\left( t,0\right) \Phi _{\alpha }\left( 0\right) $. Because $\Phi_{\alpha }\left( t\right) $ is periodic we only need to evaluate it for $t\in \lbrack 0,T] $, and from $\Phi _{\alpha }\left( t\in \lbrack 0,T]\right) $ we can directly evaluate $\Phi _{\alpha }\left( t\right) $, $\Psi _{\alpha
}\left( t\right) $ and $\Psi \left( t\right) $ for arbitrary large $t$.

A driven system that is interacting with its environment is not necessarily well described by the standard Lindblad master equation, since its dissipation process could be time-dependent due to the driving. In such cases, a rigorous approach would be to take the driving into account when deriving the master equation. This can be done in many different ways, but a one-way common approach is to derive the master equation on the Floquet basis. That approach results in the so-called Floquet-Markov master equation (see Ref. \cite{Grifoni-1998} for details).

For a summary of the derivation, the important contents for the implementation in QuTiP \cite{QuTiP1,QuTiP2} are listed below. The Floquet mode $\left\vert \phi _{\alpha }\left( t\right) \right\rangle $ refers to a full class of quasienergies defined by $\epsilon _{\alpha }+k\Omega $ for arbitrary $k$. One finds the quasienenergy difference between two Floquet modes as
\begin{equation}
\Delta _{\alpha \beta k}=\frac{\epsilon _{\alpha }-\epsilon _{\beta }}{\hbar 
}+k\Omega
\end{equation}%
For an arbitrary coupling operator $q$, the matrix elements in the Floquet basis are calculated as%
\begin{equation}
{\cal X}_{\alpha \beta k}=\frac{1}{T}\int_{0}^{T}dt\left\langle \phi _{\alpha
}\left( t\right) \right\vert q\left\vert \phi _{\beta }\left( t\right)
\right\rangle \text{.}
\end{equation}%
When the matrix elements and the spectral density $S(\omega )$ are known, the decay rate $\gamma _{\alpha \beta k}$ is defined as
\begin{equation}
\gamma _{\alpha \beta k}=2\pi \theta \left( \Delta _{\alpha \beta k}\right)
 \left\vert {\cal X}_{\alpha \beta k}\right\vert ^{2} J\left( \Delta _{\alpha \beta k}\right)                             \label{gamma0}
\end{equation}%
where $\theta(x) $ is the Heaviside function.

To describe the states of multipartite quantum systems such as a TLS coupled to an oscillator, or the multi-TLS device we expand the Hilbert space by taking the tensor product of the state vectors for each of the system components. Similarly, the operators acting on the state vectors in the combined Hilbert space (describing the coupled system) are formed by taking the tensor product of the individual operators. The state vector describing two TLS in their ground states  $ \left\vert \downarrow \right\rangle\otimes \left\vert \downarrow \right\rangle $ is formed in QuTiP~\cite{QuTiP1,QuTiP2} by taking the tensor product of the two single-TLS ground state vectors as \emph{tensor(basis(2, 0), basis(2, 0))}. This is straightforward to generalize to more TLS by adding more component state vectors in the argument list to the tensor function $\left( \left\vert \downarrow \right\rangle +\left\vert \uparrow \right\rangle \right) \otimes \left( \left\vert \downarrow \right\rangle +\left\vert \uparrow \right\rangle \right) \otimes \left\vert \downarrow \right\rangle $ which in QuTiP is \emph{tensor((basis(2, 0) + basis(2,1)).unit(), (basis(2, 0) + basis(2, 1)).unit(), basis(2, 0))}. This state is slightly more complicated, describing two TLS in a superposition between the up and down states, while the third TLS is in its ground state. To construct operators acting on an extended Hilbert space of a combined system, we similarly pass a list of operators for each component system to the tensor function. For example, to form the operator that represents the simultaneous action of the $\sigma _{x}$ operator on two TLS \emph{tensor(sigmax(), sigmax())}$\rightarrow \sigma _{x}\otimes \sigma _{x}$\emph{. } To create an operator in a combined Hilbert space that only acts on a single component, we take the tensor product of the operator acting on the subspace of interest, with the identity operators corresponding to the components that are to be unchanged. For example, the operator that represents $\sigma _{z}$ on the first TLS in a two-TLS system, while leaving the second TLS unaffected is $\sigma _{z}\otimes \hat{1}$ that is written as \emph{tensor(sigmaz(), identity(2))}.

\subsection{RWA approximation}\label{RWA}

The master equation is further simplified by the Rotating Wave Approximation (RWA), which makes the following matrix useful%
\begin{equation}
\Xi_{\alpha \beta }=\sum_{k=-\infty }^{\infty }\left[ \gamma _{\alpha \beta
k}+n_{\mathrm{th}}\left( \left\vert \Delta _{\alpha \beta k}\right\vert
\right) \right] \left( \gamma _{\alpha \beta k}+\gamma _{\alpha \beta
-k}\right)       \text{,}
\end{equation}%
where the decay rate $\gamma _{\alpha \beta k}$ is given by Eq.~(\ref{gamma0}). The density matrix of the system then evolves according to the following equations
\begin{eqnarray}
\dot{\rho}_{\alpha \alpha }\left( t\right) &=&\sum_{\nu }\left( \Xi_{\alpha
\beta }\rho _{\nu \nu }\left( t\right) -\Xi_{\nu \alpha }\rho _{\alpha \alpha
}\left( t\right) \right)  \nonumber \\
\dot{\rho}_{\alpha \beta }\left( t\right) &=&-\frac{1}{2}\sum_{\nu }\left(
\Xi_{\nu \alpha }+\Xi_{\nu \beta }\right) \rho _{\alpha \beta }\left( t\right) 
\text{ \ provided }\alpha \neq \beta
\end{eqnarray}%


\end{document}